\documentclass[review]{elsarticle}

\usepackage{lineno,hyperref}
\usepackage{amsmath}
\usepackage{float}
\usepackage{subfigure}
\usepackage{xcolor}
\usepackage{bm}

\newcommand{\fn}[1]{\left( #1 \right)}
\newcommand{\ave}[1]{\left\langle #1 \right\rangle}
\newcommand{\abs}[1]{\left| #1 \right|}
\newcommand{\epf}{\fn{1-\phi}}
\newcommand{\slip}{\abs{\ave{\mathbf{W}}}}
\newcommand{\bld}[1]{{\mathbf{#1}}} 
\makeatletter
\DeclareRobustCommand{\volume}{\text{\volumedash}V}
\newcommand{\volumedash}{%
  \makebox[0pt][l]{%
    \ooalign{$V$\cr\raisebox{0.15em}{\kern0.08em--}\cr}
  }%
}
\makeatother

\modulolinenumbers[5]

\usepackage{numcompress}\biboptions{numbers,sort&compress} 

\begin{document}

\begin{frontmatter}

\title{Particle-resolved simulation of freely evolving particle suspensions: Flow physics and modeling}

\tnotetext[mytitlenote]{This is a post-peer-review, pre-copyedit version of an article published in International Journal of Multiphase Flow. The final authenticated version is available online at: \href{https://doi.org/10.1016/j.ijmultiphaseflow.2020.103533} {https://doi.org/10.1016/j.ijmultiphaseflow.2020.103533}.\\ \copyright 2020. This manuscript version is made available under the CC-BY-NC-ND 4.0 license \href{http://creativecommons.org/licenses/by-nc-nd/4.0/}{http://creativecommons.org/licenses/by-nc-nd/4.0/}.}


\author[firstaddress,secondaddress]{Vahid Tavanashad}
\author[firstaddress,secondaddress]{Alberto Passalacqua}

\author[firstaddress,secondaddress]{Shankar Subramaniam\corref{corres_author}}
\cortext[corres_author]{Corresponding author}
\ead{shankar@iastate.edu}

\address[firstaddress]{Department of Mechanical Engineering, Iowa State University, Ames, IA 50011, USA}
\address[secondaddress]{Center for Multiphase Flow Research \& Education (CoMFRE), Iowa State University, Ames, IA 50011, USA}

\begin{abstract}
The objective of this study is to understand the dynamics of freely evolving particle suspensions over a wide range of particle-to-fluid density ratios. 
The dynamics of particle suspensions are characterized by the average momentum equation, where the dominant contribution to the average momentum transfer between particles and fluid is the average drag force.
In this study, the average drag force is quantified using particle-resolved direct numerical simulation in a canonical problem: a statistically homogeneous suspension where an imposed mean pressure gradient establishes a steady mean slip velocity between the phases.
The effects of particle velocity fluctuations, particle clustering, and mobility of particles are studied separately.
It is shown that the competing effects of these factors could decrease, increase, or keep constant the drag of freely evolving suspensions in comparison to fixed beds at different flow conditions.
It is also shown that the effects of particle clustering and particle velocity fluctuations are not independent.
Finally, a correlation for interphase drag force in terms of volume fraction, Reynolds number, and density ratio is proposed. 
Two different approaches (symbolic regression and predefined functional forms) are used to develop the drag correlation.
Since this drag correlation has been inferred from simulations of particle suspensions, it includes the effect of the motion of the particles. 
This drag correlation can be used in computational fluid dynamics simulations of particle-laden flows that solve the average two-fluid equations where the accuracy of the drag law affects the prediction of overall flow behavior.
\end{abstract}


\end{frontmatter}


\section{Introduction}

Dispersed multiphase flows are encountered when one phase in the form of bubbles, droplets, or particles is dispersed within a fluid called the carrier phase, and they include gas--solid, solid--liquid, and gas--liquid flows.
Such flows are common in both nature (e.g., solid particles or rain droplets in the air) and industry (e.g., bubble columns and fluidized bed reactors).
Understanding momentum and kinetic energy exchange between dispersed and carrier phases is central to predicting the behavior of many multiphase flows.
Although we can use both experimental and numerical studies to explore the interaction between carrier and dispersed phases, particle-resolved direct numerical simulation (PR-DNS) has proven to be a useful tool for understanding flow physics and model development \citep{tenneti_arfm_2014,tryggvason_review_2013}.
However, PR-DNS of industrial multiphase processes in realistic geometries at scale is not feasible, even on today's supercomputers, due to its computational cost.

On the other hand, multiphase computational fluid dynamics (CFD) simulations that solve the averaged equations of multiphase flow are increasingly being used as an efficient alternative for design optimization because experiments are often costly and time-consuming. 
CFD simulations of multiphase flow are based on either the Eulerian--Lagrangian (EL) or the Eulerian--Eulerian (EE) two-fluid approach. 
In the EE method, each phase is treated as a continuous medium interpenetrating the other phase and is represented by macroscopic conservation equations, which are valid throughout the entire flow domain.
The averaging process results in unclosed terms that represent interphase interactions and need to be modeled. 
For instance, the mean momentum conservation equation in the particle phase requires closure of the mean drag force. 
This closure for the mean drag force is popularly known as a \textit{drag law} and is typically obtained from a combination of theoretical, experimental, and computational studies.

In the EL approach, the trajectory of each particle is tracked in response to collisional and hydrodynamic forces, while the carrier flow is represented in a Eulerian frame.
The particles can be considered \textit{point particles} if their diameter is smaller than the smallest scales of fluid motion and in this case, we can use a grid size larger than the size of the particles which means the flow field on the particle surface is not resolved.
The majority of EL simulations to date have been using the point-particle approach.
However, in recent years, EL methods are being extended to finite-size particles, whose diameter is comparable to the mesh spacing, using volume-filtering \citep{capecelatro_2013}.
In the case of finite-size EL methods, there are still outstanding questions as to how to couple the dispersed and carrier phases \citep{shankar_bala_2018}. 
Therefore, the interaction between particles and the surrounding flow, which is typically referred to as the drag correlation must be modeled using empirical relations or PR-DNS.

An accurate drag correlation for the representation of the average interphase momentum transfer term is essential to perform predictive CFD simulations.
However, it should be mentioned that there are different averaging approaches for EE methods including time--averaging \citep{ishii_book}, volume--averaging \citep{anderson_1967,jackson_1997}, ensemble--averaging \citep{drew_1983,drew_book,pai_subramaniam_2009}.
The volume--averaging approach by \citet{anderson_1967} also involved filtering the fluid and particle fields.
\citet{capecelatro_2013} extended this filtering approach to EL method and developed the volume-filtered Euler-Lagrange (VFEL) method.

Although these averaging methods look similar, they are fundamentally different, and there has been a discussion on their validity and connection in the literature \citep{joseph_1990}.
Recently, \citet{shankar_2020} showed the differences between the volume--averaging and ensemble--averaging, employing a simple example based on the mass conservation equation.
Consistency of the PR-DNS and EE approaches requires that the method used to calculate the surface force density in the PR-DNS should be consistent
with the definition of the average interphase momentum transfer term. 
Otherwise, the drag correlation inferred from PR-DNS may not be consistent with the EE equations that arise from the two-fluid theory \citep{tenneti_ijmf_2011}.
A detailed discussion on the connection of PR-DNS equations and ensemble-averaged two-fluid equations is presented by \citep{tenneti_ijmf_2011}.

Several researchers have studied the interphase momentum transfer (drag exchange) between phases in particle-laden flows. 
This is usually done in an idealized canonical flow problem in which the dispersed phase consists of monodisperse spherical objects which are fixed and distributed homogeneously in a periodic domain.
Fixed beds are a good approximation for gas--solid flows with high inertia particles.
This special case is well-studied and several drag correlations are proposed in the literature \citep{hill_koch_2001a,hill_koch_2001b,vanderHoef_2005,beetstra_2007,tenneti_ijmf_2011, rong_2013,zaidi_2014,bonger_2015,tang_2015,kravets_2019}.
These studies have also been extended to bidisperse particles \citep{vanderHoef_2005,beetstra_2007,yin_sundaresan_2009a,yin_sundaresan_2009b, mehrabadi_ijmf_2016} as well as clustered \citep{mehrabadi_ces_2016} or inhomogeneous \citep{wang_2011,zhou_2014} configuration of particles.

As an improvement of fixed bed simulations, simulating stationary particles with an assigned non-zero velocity has been performed to investigate the effect of fluctuating particle acceleration on particle velocity fluctuations \citep{tenneti_pt_2010} and the effects of particle velocity fluctuations on interphase drag \citep{wylie_koch_ladd_2003,huang_2017,bala_PFR_2020} and heat transfer \citep{huang_2019} in gas--solid flows.

Particle-resolved simulations of freely evolving suspensions of gas--solid and solid--liquid flows \citep{tenneti_pt_2010,kriebitzsch_2013, zhou_2014,luo_2016,tang_free_2016,rubinstein_2016,rubinstein_2017, zaidi_2018} 
have also been performed in the past years with a focus on studying the interphase drag force.
Although the effects of particle velocity fluctuations, clustering, or mobility on mean drag have been investigated in these studies, none of them present a complete description of the effects of these three factors.
These works only explain the change of drag compared to fixed beds by considering one factor at a time.
The only exception is the work by \citet{rubinstein_2017}, who consider both particle mobility and clustering but only for low-Reynolds number flows.

Most recently, \citet{tavana_ijmf_2020} proposed a PR-DNS-based drag law for buoyant particle suspensions which are a good approximation to spherical bubbles in contaminated liquid \citep{clift_book,magnaudet_2000,takagi_Matsumoto_2011}.
They showed that with proper scaling, the drag of buoyant particles is comparable with the drag of bubbles in clean liquid \citep{gillissen_2011,roghair_2013}. 

In this work, we have performed PR-DNS for a wide range of density ratio to cover both heavy and light particles.
Then we have studied the effects of the factors mentioned above (particle clustering, particle velocity fluctuations, and particle mobility) separately and also altogether in particle-laden flows.
Finally, an improved drag correlation is proposed, which can be used for calculating the drag force in EE and EL simulations of particles with different densities.
In developing our correlation, we have used two different approaches: symbolic regression and a predefined functional form.
We have also discussed which variables the drag correlation should depend on to account for particle motion.
In the following, we have summarized the discussion on the effects of particle clustering, particle velocity fluctuations, and mobility of particles on the mean drag in dispersed multiphase flows from the literature.

\textit{Particle clustering}: The emergence of clustering in the simulation of freely evolving suspensions of solid particles or bubbles has already been reported in the literature.
Prior works have shown that nearly spherical bubbles form clusters and generate horizontal planes of bubbles, known as rafts, perpendicular to the flow direction in bubbly flows, which increases the drag force \citep{bunner_tryggvason_2002a,esmaeeli_tryggvason_2005,yin_koch_2008,roghair_2013}.
\citet{yin_koch_2008} compared the microstructure of particle and bubble suspensions at intermediate Reynolds numbers $\fn{Re = 5.4, 20}$ and volume fractions $\fn{\phi <0.25}$ and showed that horizontal clustering occurs in both systems but it is more significant for bubble suspensions.
On the other hand, vertically elongated columnar particle clusters are observed in dilute systems at high Reynolds number in gas--solid flows, which reduce the average drag force \citep{uhlmann_doychev_2014,zaidi_2018}.
Moreover, \citet{wang_2011} and \citet{zhou_2014} have found from the simulation of an inhomogeneous fixed bed that the drag force depends on both the direction and magnitude of the particle volume fraction gradient, with volume fraction gradients in the direction of the mean slip velocity causing an increase in drag, and volume fraction gradients perpendicular to the slip velocity causing a decrease in drag.
It is also known that isotropic clustering in the fixed assembly of particles always decreases the drag \citep{wang_2011,mehrabadi_ces_2016}.
The importance of considering the particle structure in modeling drag has also been the topic of a recent review paper by \citet{sundaresan_review_2018}.

\textit{Particle velocity fluctuations}: The mean relative motion of particles/bubbles is responsible for the generation of fluid velocity fluctuations, which in turn modify velocity fluctuations in particles/bubbles and the mean relative motion (or drag force) between phases \citep{mehrabadi_pf_2017,risso_arfm_2017}.
In prior works, it is shown that particle velocity fluctuations act as a source for an increase in the drag of gas--solid flows \citep{wylie_koch_ladd_2003,tenneti_pt_2010,tang_free_2016,wang_2017}. 
Most recently, \citet{tavana_acta_2019} studied particle suspensions for a wide range of density ratio $\fn{0.001 \le \rho_p/\rho_f \le 1000}$ at Reynolds number $20$.
They showed that for this Reynolds number, drag does not change significantly with density ratio even though particle and fluid velocity fluctuations increase with decreasing density ratio.
The present work extends the range of that study and examines whether this trend persists at all Reynolds number in the range $10$ to $100$.

\textit{Particle Mobility}: The ability of the particles to translate and rotate due to the effects of the surrounding fluid can decrease the drag on particles \citep{rubinstein_2016}.
In fact, light particles follow the streamlines of fluid, and for heavier ones, they continue moving on their initial trajectory.

\section{Numerical method}
\label{sec:method}
The PR-DNS approach used in this work is based on the discrete-time direct forcing immersed boundary method of \citet{jamalphd} and is called the particle-resolved uncontaminated-fluid reconcilable immersed boundary method (PUReIBM). 
The PUReIBM methodology is explained in detail and validated for simulating fixed beds \citep{garg_book,tenneti_ijmf_2011}, gas--solid flows \citep{tenneti_pt_2010,mehrabadi_jfm_2015,tenneti_jfm_2016}, and buoyant particles \citep{tavana_ijmf_2020}. 
Here, the main features of this method are presented.

In PUReIBM, the continuity and the Navier--Stokes equations are solved for the fluid phase on a uniform Cartesian grid with the Crank--Nicolson scheme for the viscous terms and an Adams--Bashforth scheme for the convective terms.
The boundary conditions on the fluid velocity at the particle interface (no-slip and no-penetration) are imposed via a source term (immersed boundary force) in the Navier-Stokes equations.

The motion of each particle in PUReIBM is evolved by updating its position and translational and rotational velocities, according to Newton's second law.
A soft-sphere collision \citep{cundall_dem} model is used to capture particle-particle interactions. 
Particles are allowed to overlap during a collision, and the contact mechanics between the overlapping particles are modeled by a spring in the normal direction (elastic collisions).
The spring causes the colliding particles to rebound.
The particles are assumed to be frictionless during collisions.
This implies that the tangential component of the contact force is zero.

To stabilize the simulations for buoyant particles, the virtual force stabilization technique introduced by \citet{schwarz_Kempe_2015} is utilized, which is extended for simulation of dense suspensions \citep{tavana_ijmf_2020}.
The simulation of buoyant particles in this work can also be considered as an approximation to bubbly flows when the presence of surfactants contaminates the surrounding fluid, and the no-slip velocity boundary condition is valid at the interface of bubbles \citep{clift_book,magnaudet_2000,takagi_Matsumoto_2011}.

\section{Simulation setup}
\label{sec:setup}
In this study, simulations are performed in a cubic domain with periodic boundary conditions.
A constant mean pressure gradient is imposed in the $x$-direction that accelerates the particles and the fluid.
Both the mean fluid velocity and the mean particle velocity increase; however, their difference---the mean slip velocity---reaches a statistically stationary state.
The magnitude of the mean pressure gradient depends on three parameters: the dispersed phase volume fraction $\phi$, the particle-to-fluid density ratio $\rho_p/\rho_f$, and a Reynolds number defined as:
\begin{equation}
\label{eq:Re_m}
Re_m = \frac{\rho_f \epf \slip {d_p}}{\mu_f}, \nonumber
\end{equation}
where $d_p$ is the particle diameter, $\mu_f$ is the dynamic viscosity of the fluid phase, and $\ave{\mathbf{W}}$ is the mean slip velocity between the particles and the fluid.

For each set of physical parameters, five independent realizations (corresponding to a specified initial particle configuration) are simulated in this study.
The initial positions of the particles are obtained following elastic collisions (in the absence of interstitial fluid), starting from a lattice arrangement with a Maxwellian velocity distribution.
All the mean quantities in the fluid phase are computed by first volume-averaging the flow variable for one realization and, subsequently, ensemble-averaging over different particle realizations.
The mean quantities in the dispersed phase are computed by averaging over all particles and then ensemble-averaging over different particle configurations.

Although the primary goal of this work is the simulation of freely evolving suspensions, in order to separate the effects of particle velocity fluctuations, clustering and mobility on the drag of moving particles, we have performed, in total, five different types of simulations which are summarized in Table \ref{tab:setup}.
The base simulations are for fixed assemblies of homogeneous particles (Case 1: Homogeneous Fixed).
To show the effect of particle clustering on the drag force, we use the configuration of particles from the simulation of freely evolving suspensions after it reaches a statistically stationary state (steady values of second moments of particle and fluid velocities) and simulate it as a fixed bed of stationary particles (Case 2: Clustered Fixed).
Assigning a random fluctuating velocity sampled from moving particle simulations to each particle in the homogeneous fixed bed allows considering the pure effect of particle velocity fluctuations on the hydrodynamic forces (Case 3: Homogeneous Snapshot).
If a random fluctuating velocity sampled from moving particle simulations is assigned to each particle in the clustered fixed bed, then the fixed bed simulation can be considered as an instantaneous snapshot of a freely evolving suspension (Case 4: Clustered Snapshot).
Of course, in a freely evolving suspension, the dynamic response of the particles to the hydrodynamic forces will affect the particle velocity and position, and this is not captured by the snapshot simulation. We have considered this effect by simulating freely evolving suspensions (Case 5: Freely Evolving).

\begin{table} [H]
\centering
\addtolength{\leftskip} {-1cm}
\caption{The different types of simulations classified as cases that are considered in this study.}
\label{tab:setup} 
\begin{tabular}{c c c c c} 
\hline\noalign{\smallskip}
Case & Name & particle configuration & mobility & particle velocity  \\
\noalign{\smallskip}
\hline\noalign{\smallskip}
$1$ & Homogeneous Fixed & homogeneous & stationary & zero \\ 
$2$ & Clustered Fixed & clustered & stationary & zero \\ 
$3$ & Homogeneous Snapshot & homogeneous & stationary & non-zero \\ 
$4$ & Clustered Snapshot & clustered & stationary & non-zero \\ 
$5$ & Freely Evolving & clustered & moving & non-zero \\
\noalign{\smallskip}
\hline
\end{tabular}
\end{table}

The salient numerical and physical parameters used in the simulations are reported in Table \ref{tab:num_param}.

\begin{table} [H]
\centering
\caption{The numerical and physical parameters of the simulations: volume fraction of particles $\phi$, the number of grid cells across the diameter of a particle $d_p/\Delta x$ (numbers before the "/" correspond to $Re_m<100$ while numbers after the "/" correspond to $Re_m=100$), the ratio of the length of the box to the particle diameter $L/d_p$, number of particles $N_p$, Reynolds number $Re_m$, and particle-to-fluid density ratio $\rho_p/\rho_f$.}
\label{tab:num_param} 
\begin{tabular}{c c c c c c c} 
\hline\noalign{\smallskip}
$\phi$ & $d_p/\Delta x$ & $L/d_p$ & $N_p$ & $Re_m$ & $\rho_p/\rho_f$\\
\noalign{\smallskip}
\hline\noalign{\smallskip}
$0.1$ & $20/30$ & $10.16$ & $200$ & $10,20,50,100$ & $0.01,0.1,10,100$ \\
$0.2$ & $20/30$ & $8.06$ & $200$ & $10,20,50,100$ & $0.01,0.1,10,100$\\
$0.3$ & $30/40$ & $7.05$ & $200$ & $10,20,50,100$ & $0.01,0.1,10,100$\\
$0.4$ & $30/40$ & $6.4$ & $200$ & $10,20,50,100$ & $0.01,0.1,10,100$\\
\noalign{\smallskip}
\hline
\end{tabular}
\end{table}

\section{Results and discussion}
\label{sec:results}
\subsection{Mean drag force}
\label{sec:mean_drag}
\begin{figure} [H]
\begin{centering}
\subfigure[]{ \includegraphics[clip, width=57mm]{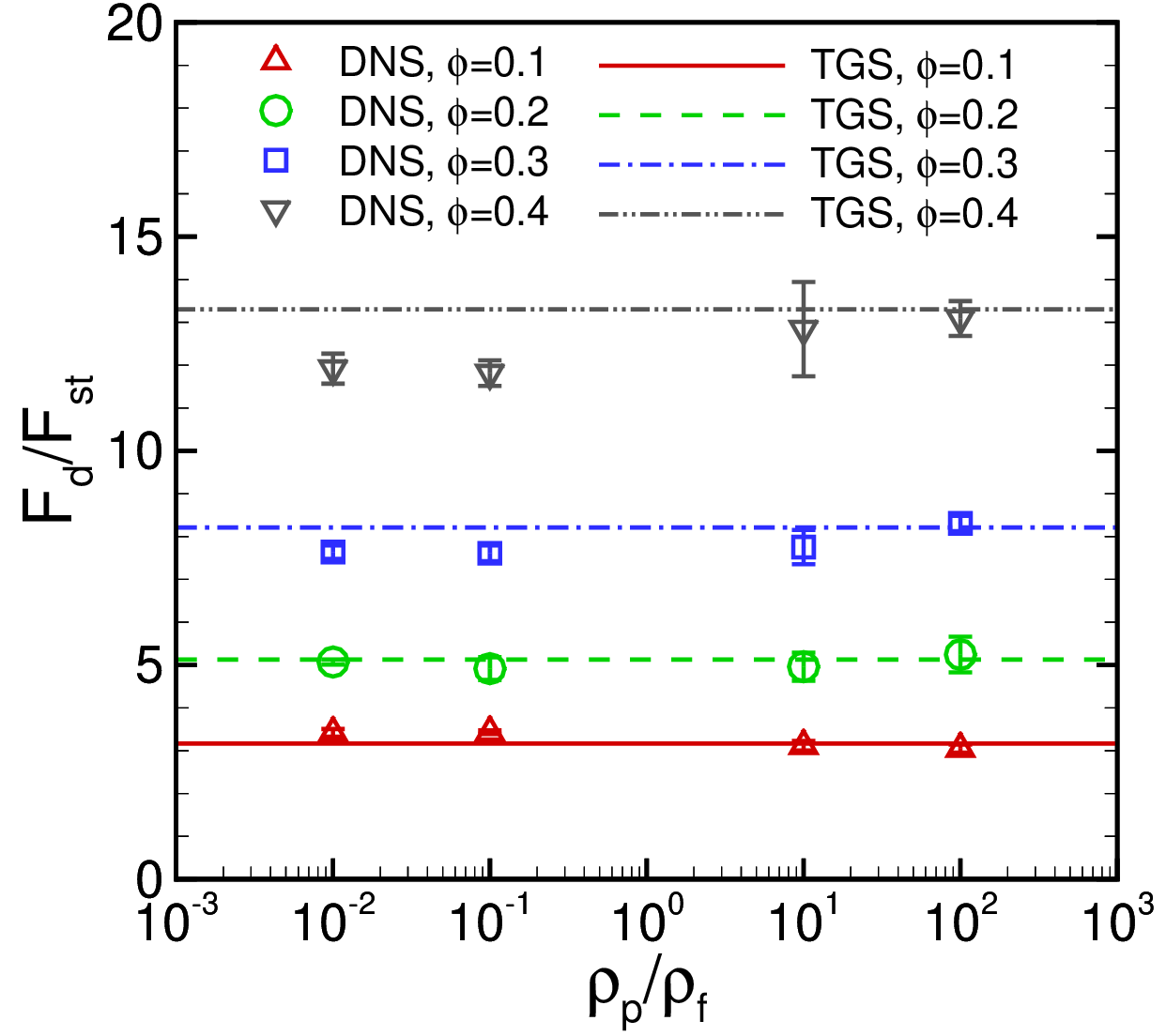} \label{fig:drag1_TGS}}
\subfigure[]{ \includegraphics[clip, width=57mm]{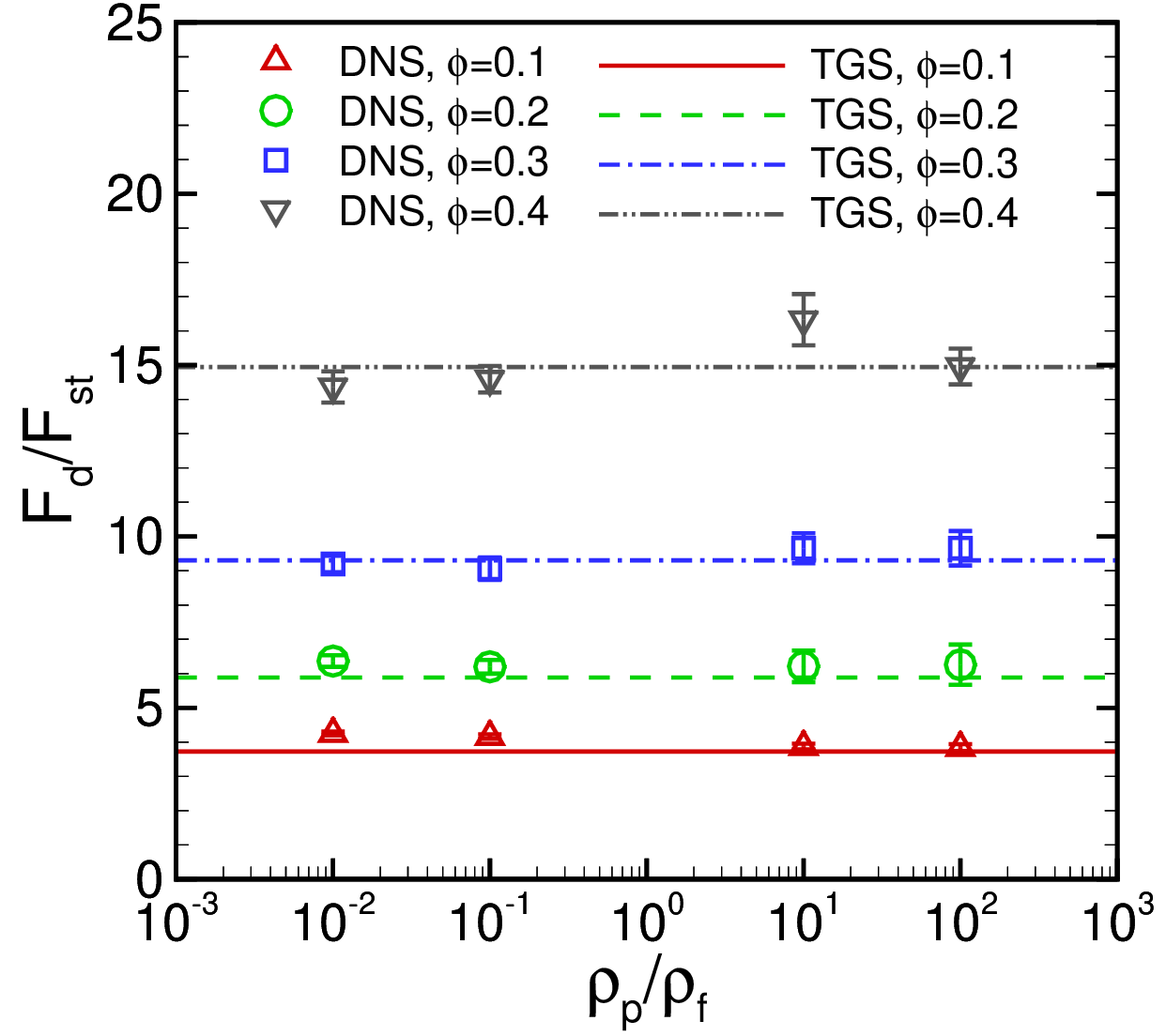} \label{fig:drag2_TGS}}
\subfigure[]{ \includegraphics[clip, width=57mm]{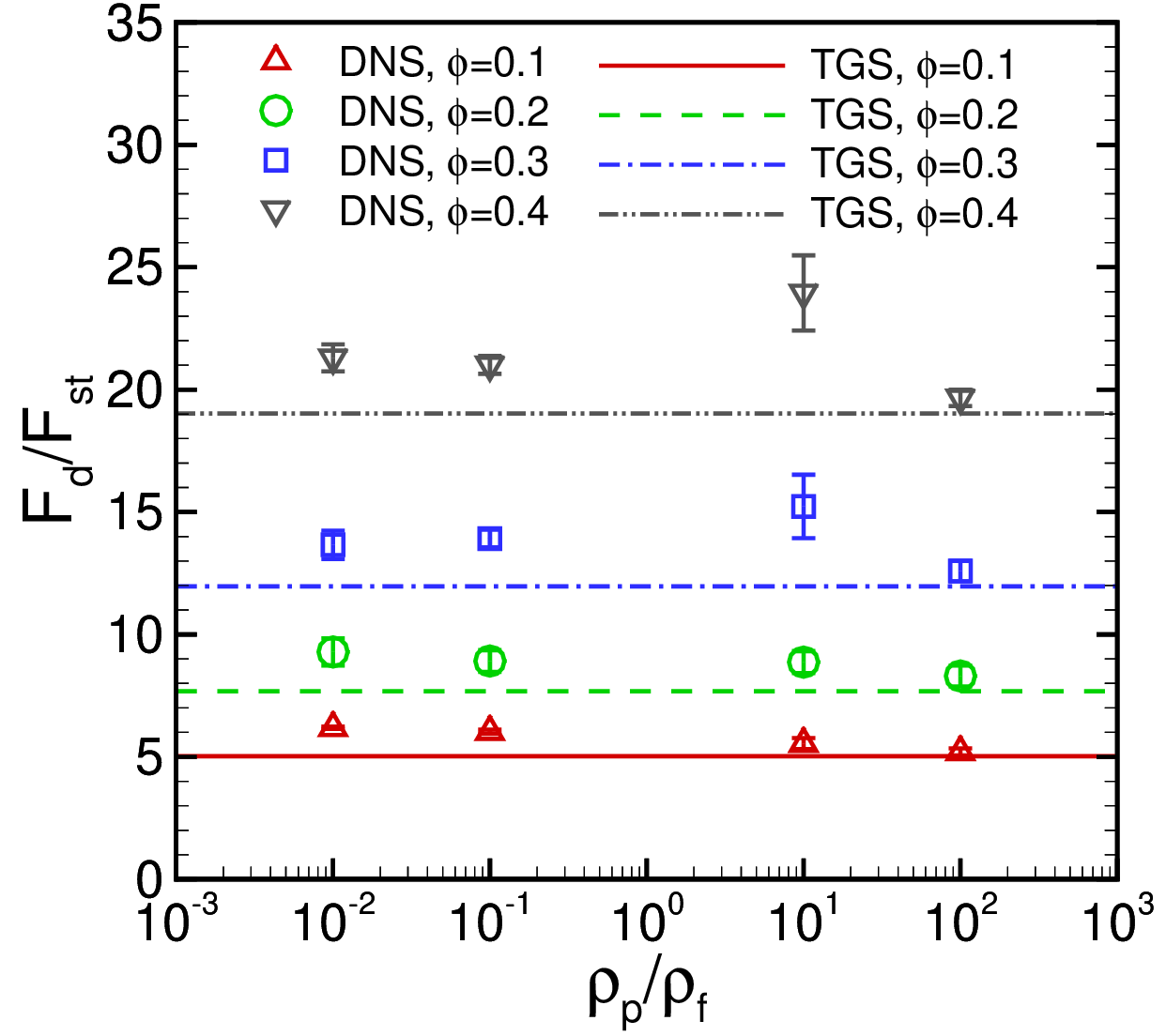} \label{fig:drag3_TGS}}
\subfigure[]{ \includegraphics[clip, width=57mm]{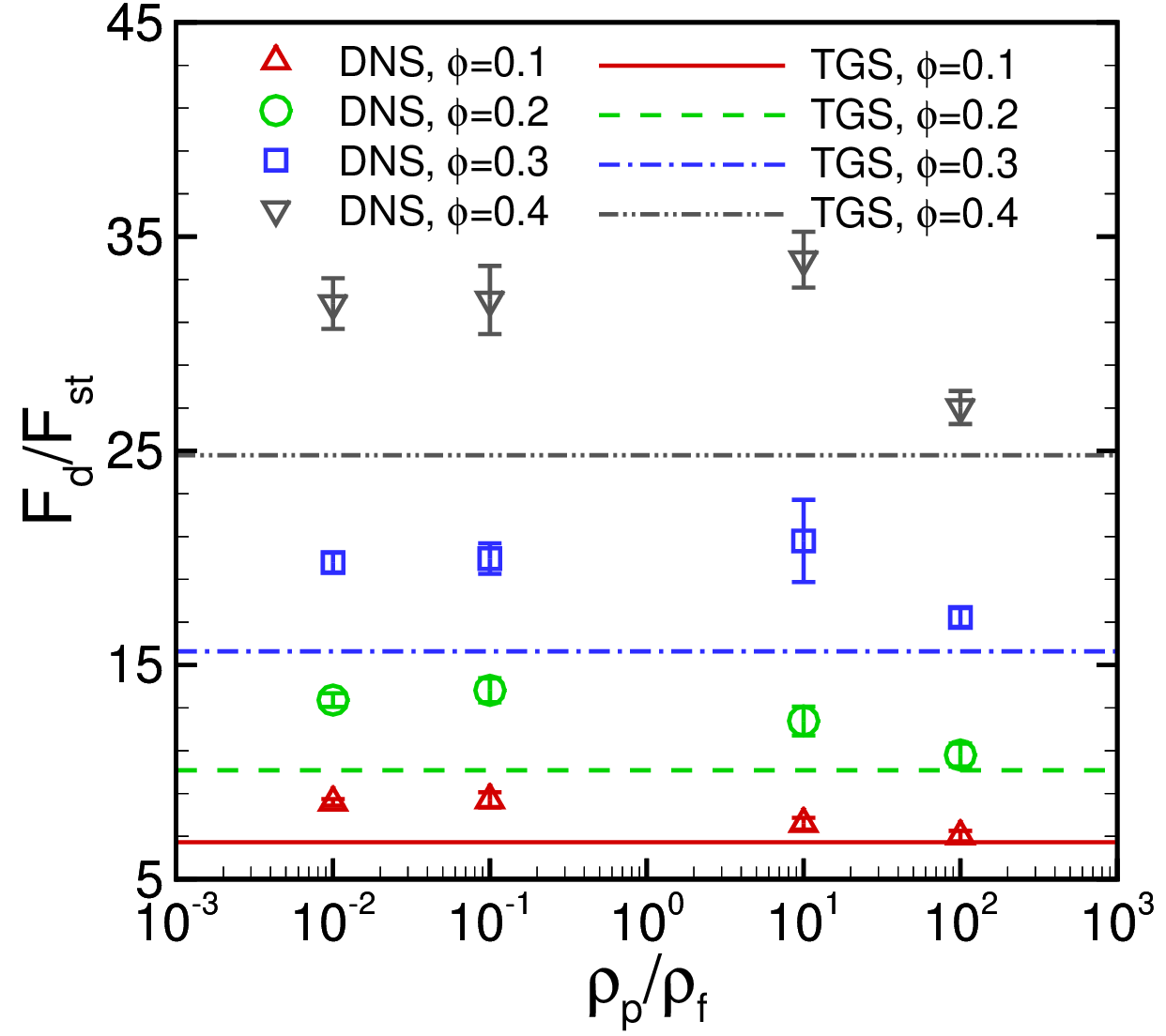} \label{fig:drag4_TGS}}
\caption{Non-dimensional drag force $F= F_d/F_{st}$, as a function of ${\rho_p/\rho_f}$ for different $Re_m$ and $\phi$.
The drag correlation by \protect\citet{tenneti_ijmf_2011}, denoted by TGS, for fixed beds, is also shown for comparison.
Symbols show PR-DNS data, and lines represent the correlation.
The error bars represent $95 \%$ confidence intervals obtained from five independent realizations for each case.
\subref{fig:drag1_TGS} $Re_m=10$.
\subref{fig:drag2_TGS} $Re_m=20$.
\subref{fig:drag3_TGS} $Re_m=50$.
\subref{fig:drag4_TGS} $Re_m=100$.}
\label{fig:drag_TGS}
\end{centering}
\end{figure}

Figure \ref{fig:drag_TGS} shows the non-dimensional drag force $F = F_d/F_{st}$ (see \ref{sec:appA} for details on $F_d$) as a function of density ratio for different Reynolds numbers and volume fractions in freely evolving suspensions compared to the drag correlation by \citet{tenneti_ijmf_2011} for fixed beds.
Here $F_{d}$ is the dimensional average drag on each particle and $F_{st}=3 \pi \mu_f \epf d_p \slip $ is Stokes drag.
Note that $F_{d}$ does not include the effect of the mean pressure gradient. 
The relation between this drag force and total fluid-particle force, which includes the effect of the mean pressure gradient, is explained in \ref{sec:appA}.

This figure shows that depending on the flow conditions, the drag of a freely evolving suspension could be smaller than, greater than, or the same as in its fixed bed counterpart.
To further examine the difference between mean drag in moving particles and fixed beds, we have studied the effects of particle velocity fluctuations, clustering, and mobility in the following subsections.

Table \ref{tab:delta_F} summarizes the main conclusions from this study, which are supported by providing the results in the following subsections.
It is shown that the presence of the clustering in particle configuration increases the drag by generating horizontal clusters for the range of parameters studied in this work.
The particle velocity fluctuations for the homogeneous configuration also increases the drag.
However, the effect of particle velocity fluctuations on the drag for the clustered case does not have a specific trend and could increase or decrease the drag force.
This indicates that the effects of clustering and particle velocity fluctuations are not independent.
Finally, the mobility of particles decreases the drag.

\begin{table} [H]
\centering
\caption{The effects of different factors on drag.}
\label{tab:delta_F} 
\begin{tabular}{c c c} 
\hline\noalign{\smallskip}
Parameter & Effect on the drag & Quantification \\
\hline\noalign{\smallskip}
clustering & increase & $ F_{\textrm{fixed}}^{\textrm{clustered}} - F_{\textrm{fixed}}^{\textrm{homogeneous}} $ \\ 
\hline\noalign{\smallskip}
\begin{tabular}{c} particle velocity fluctuations\\(homogeneous) \end{tabular} & increase & $ F_{\textrm{snapshot}}^{\textrm{homogeneous}} - F_{\textrm{fixed}}^{\textrm{homogeneous}} $ \\
\hline\noalign{\smallskip}
\begin{tabular}{c} particle velocity fluctuations\\(clustered) \end{tabular} & increase/decrease & $ F_{\textrm{snapshot}}^{\textrm{clustered}} -F_{\textrm{fixed}}^{\textrm{clustered}} $ \\
\hline\noalign{\smallskip}
mobility & decrease & $ F_{\textrm{moving}}^{\textrm{clustered}} - F_{\textrm{snapshot}}^{\textrm{clustered}} $ \\
\noalign{\smallskip}
\hline
\end{tabular}
\end{table}

\subsection{Effect of particle clustering}
\label{sec:cluster}
\begin{figure} [H]
\begin{centering}
\includegraphics[scale=0.4]{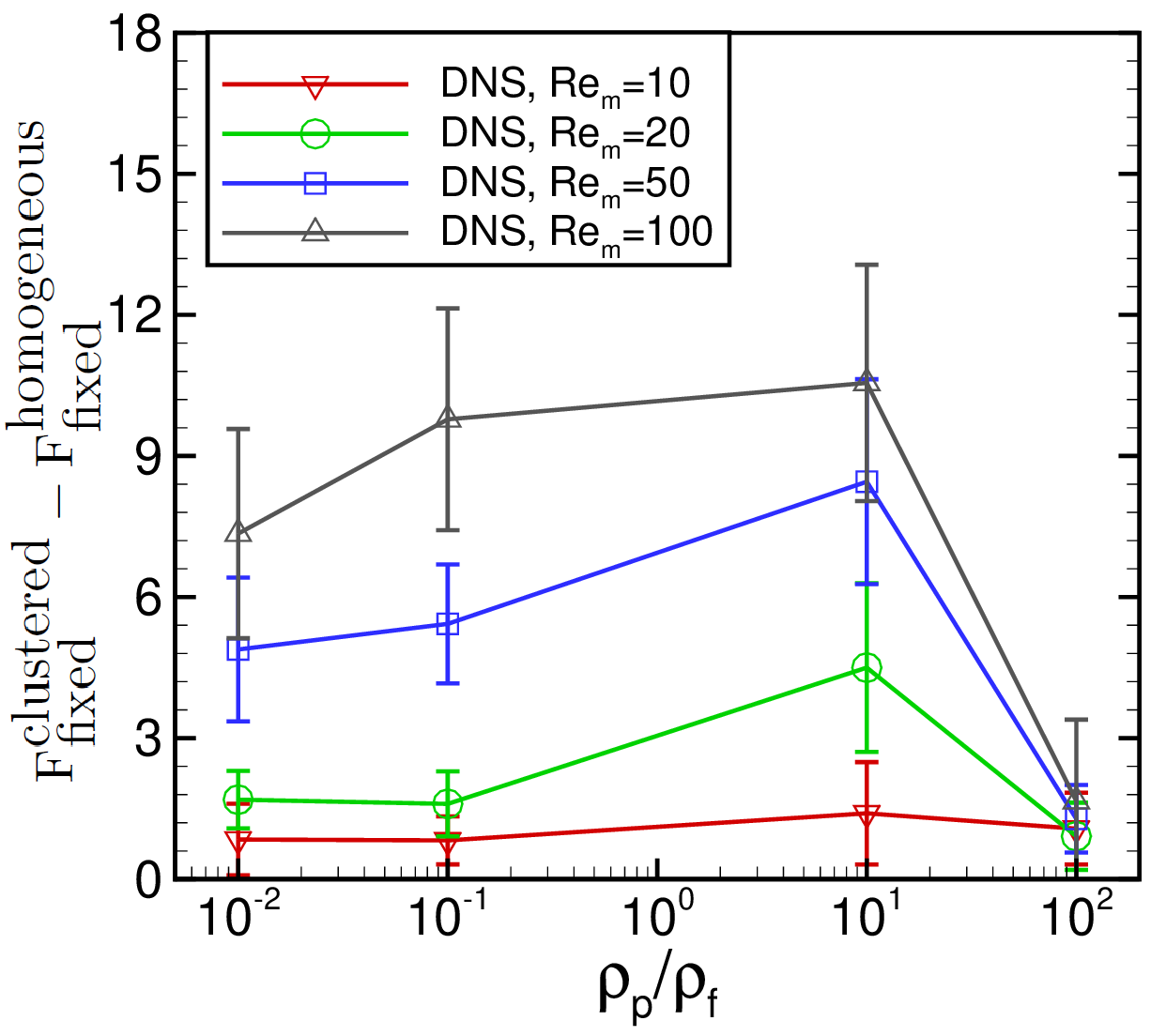}
\caption{Drag difference scaled with Stokes drag between homogeneous and clustered fixed bed (Cases 1 and 2) as a function of $\rho_p/\rho_f$ for different $Re_m$ and $\phi=0.4$.
The error bars represent $95 \%$ confidence intervals obtained from five independent realizations for each case.}
\label{fig:cluster}
\end{centering}
\end{figure}

Figure \ref{fig:cluster} shows the change in drag of clustered fixed particle configurations (Case 2) in comparison to homogeneous fixed assemblies (Case 1).
It shows that the drag of clustered particles increases in comparison with homogeneous cases, which suggests the presence of horizontal clustering or rafts.

To quantify particle clustering in our simulations and explain its connection to the increase in drag, we calculate the radial and angular pair distribution functions which are defined as \citep{bunner_tryggvason_2003}:
\begin{align}
g \fn{r} & = \frac{\volume_{sys}}{N_p \fn{N_p-1}} \frac{1}{\Delta \volume \fn{r}} \sum_{i=1}^{N_p} \sum_{j=1,i \neq j}^{N_p} \delta \fn{r-\frac{1}{2} \Delta r \le R < r+\frac{1}{2} \Delta r}, \nonumber \\
g \fn{\theta} & = \frac{\volume_{sys}}{N_p \fn{N_p-1}} \frac{1}{\Delta \volume \fn{\theta}} \sum_{i=1}^{N_p} \sum_{j=1,i \neq j}^{N_p} \delta \fn{\theta-\frac{1}{2} \Delta \theta \le \Theta < \theta + \frac{1}{2} \Delta \theta}, \nonumber
\end{align}
where $\volume_{sys}$ is the volume of the system, ${\Delta \volume \fn{r}}$ is the volume of a spherical shell between $ r - \Delta r/2$ and $ r + \Delta r/2$, ${\Delta \volume \fn{\theta}}$ is the volume of the spherical sector of radius $r^{*}$ contained within the angles $\theta-\frac{1}{2} \Delta \theta$ and $\theta+\frac{1}{2} \Delta \theta$, $R$ is the distance between the centroids of particles $i$ and $j$, $\Theta$ is the angle between the flow direction and the centerline of particles $i$ and $j$, and $\delta \fn{ \cdot }$ is equal to one if the condition in parentheses is true and zero otherwise.

\begin{figure} [H]
\begin{centering}
\subfigure[]{ \includegraphics[clip, width=57mm]{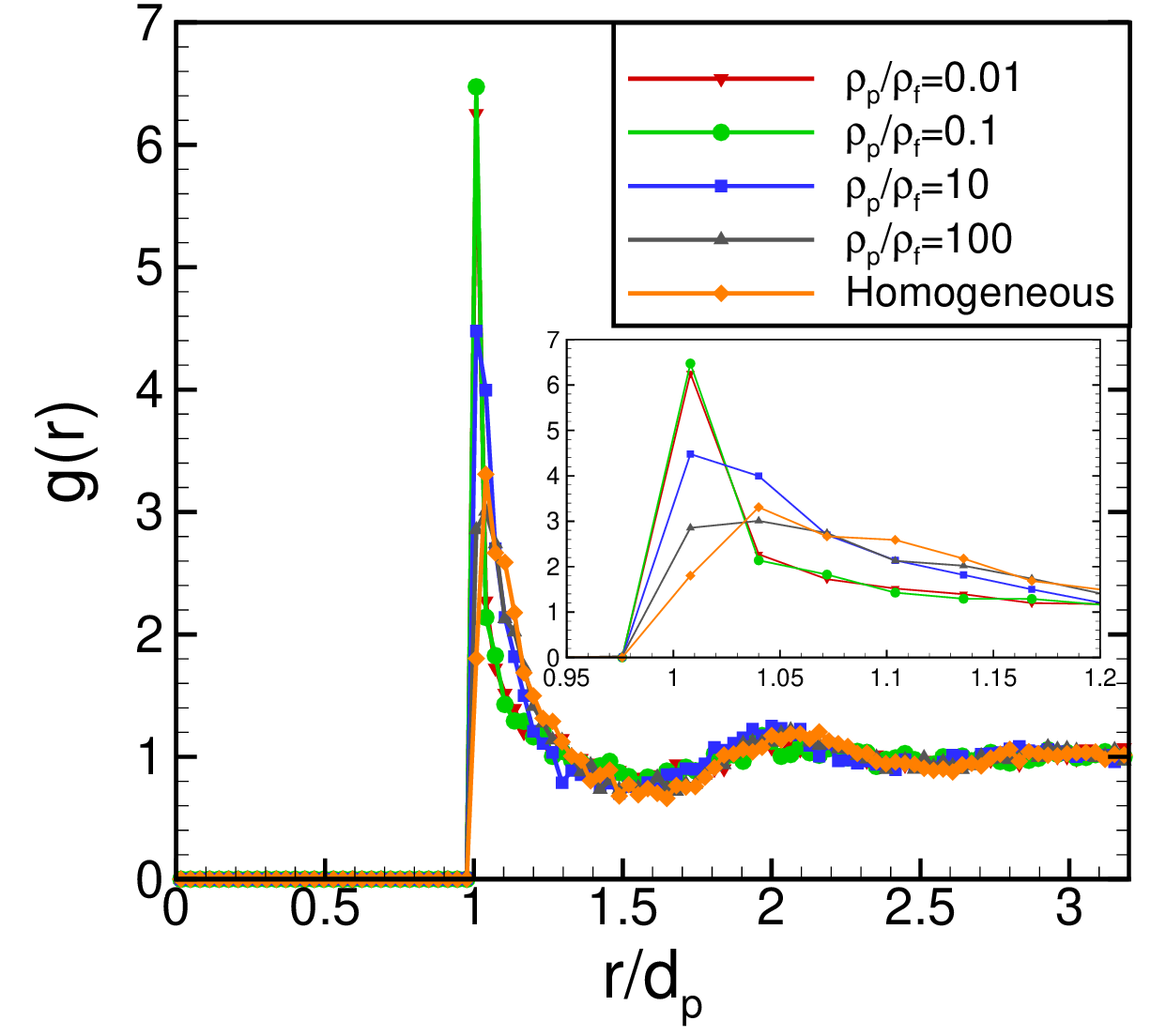} \label{fig:rdf}}
\subfigure[]{ \includegraphics[clip, width=57mm]{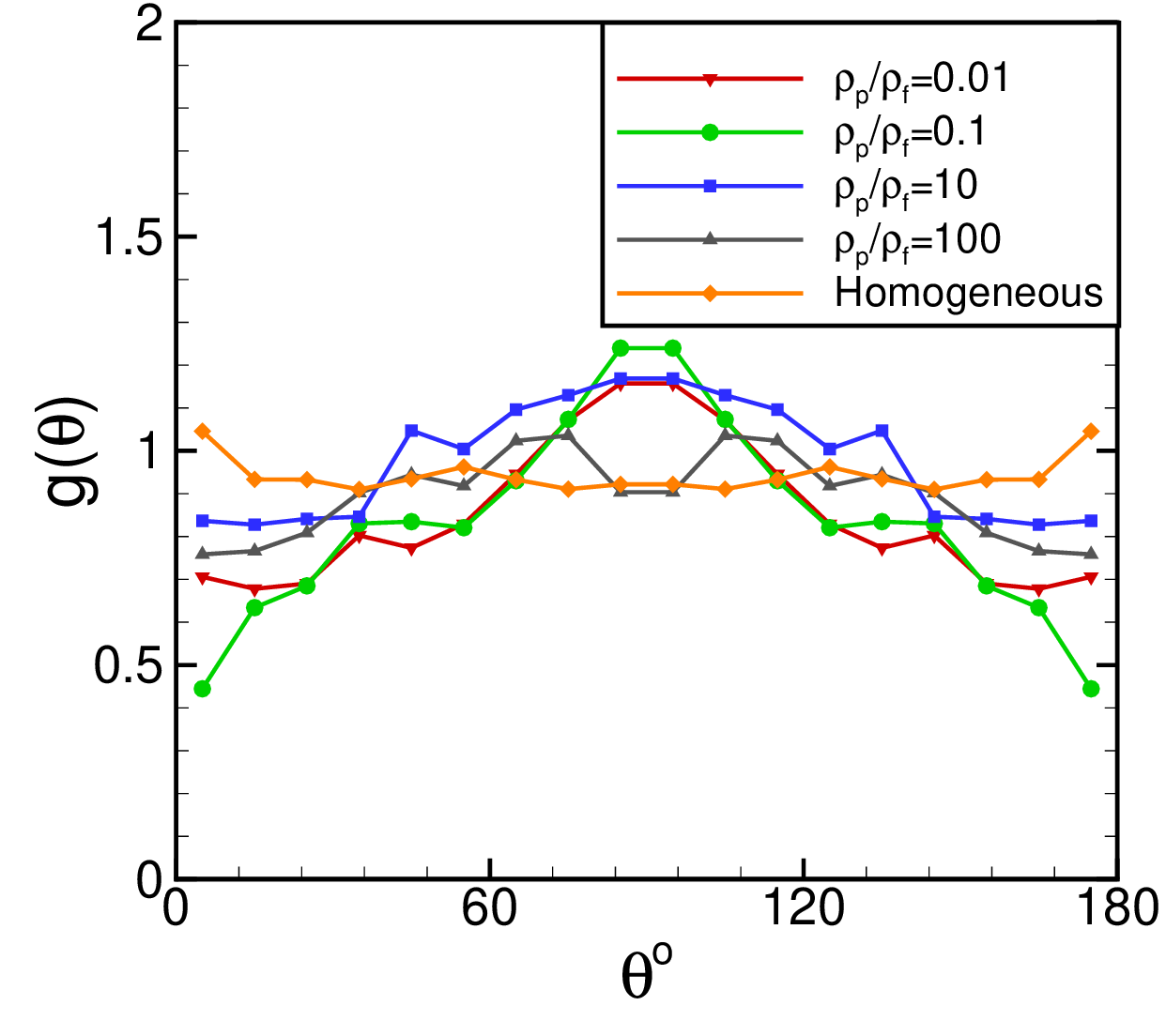} \label{fig:adf}}
\caption{Pair distribution functions for fixed bed (homogeneous) and freely evolving cases with different density ratios at $Re_m=50$ and $\phi=0.4$.
\subref{fig:rdf} Radial pair distribution function.
\subref{fig:adf} Angular pair distribution function with $ r^{*} = 1.25 d_p$.}
\label{fig:pdf}
\end{centering}
\end{figure}

Figure \ref{fig:rdf} shows the radial pair distribution function for different density ratios and $Re_m=50$.
The higher peak at low-density ratios indicates larger clusters for buoyant particles and explains the larger difference in the drag seen in Fig. \ref{fig:cluster}.
Such a dependence of the peak on density ratio is observed at all $Re_m$, but not shown here.
As $r/d_p$ becomes larger, $g\fn{r}$ goes to one which means at larger scales the distribution of particles in all cases is uniform and particle clustering is happening locally.
Figure \ref{fig:adf} shows the angular pair distribution function for the same cases when $r^{*}= 1.25 d_p$ is chosen.
The peak close to $\theta =  90^{\circ}$ is a sign of horizontal clustering and explains the increase of drag ($\theta=0^{\circ}$ is defined in the flow direction).
If we use a larger value for $r^{*}$, $g \fn{\theta}$ becomes almost uniform and equal to one for all cases, which again shows that particle positions are uncorrelated at larger scales.
This observation is consistent with the results for bubbly flows reported in \citet{bunner_tryggvason_2002a,bunner_tryggvason_2003}.

Although Fig. \ref{fig:cluster} only shows results for $\phi=0.4$, the increase of drag due to clustering is seen for all cases studied in this work.
However, as we mentioned in the Introduction, a decrease in drag due to vertical clustering is reported in the simulation of moving particles in dilute systems $\fn{\phi < 0.01}$ at high Reynolds number $\fn{ Re > 200}$ \citep{uhlmann_doychev_2014,zaidi_2018}.
This is because columnar clusters reduce drag but rafts (horizontal) clusters increase drag. 

\subsection{Effect of particle velocity fluctuations}
\label{sec:fluctuation}
To characterize the effect of particle velocity fluctuations, it is useful to define a Reynolds number based on the granular temperature as:
\begin{equation}
\label{eq:Re_T}
Re_T = \frac{\rho_f T^{1/2} {d_p}}{\mu_f}, \nonumber
\end{equation}
where ${T= \ave{\mathbf{v}''\cdot \mathbf{v}''}/3}$ is the granular temperature and $\mathbf{v}''= \mathbf{v} - \ave{\mathbf{v}}$ denotes the fluctuation in the particle velocity $\mathbf{v}$ with respect to the mean particle velocity $\ave{\mathbf{v}}$.

\begin{figure} [H]
\begin{centering}
\includegraphics[scale=0.4]{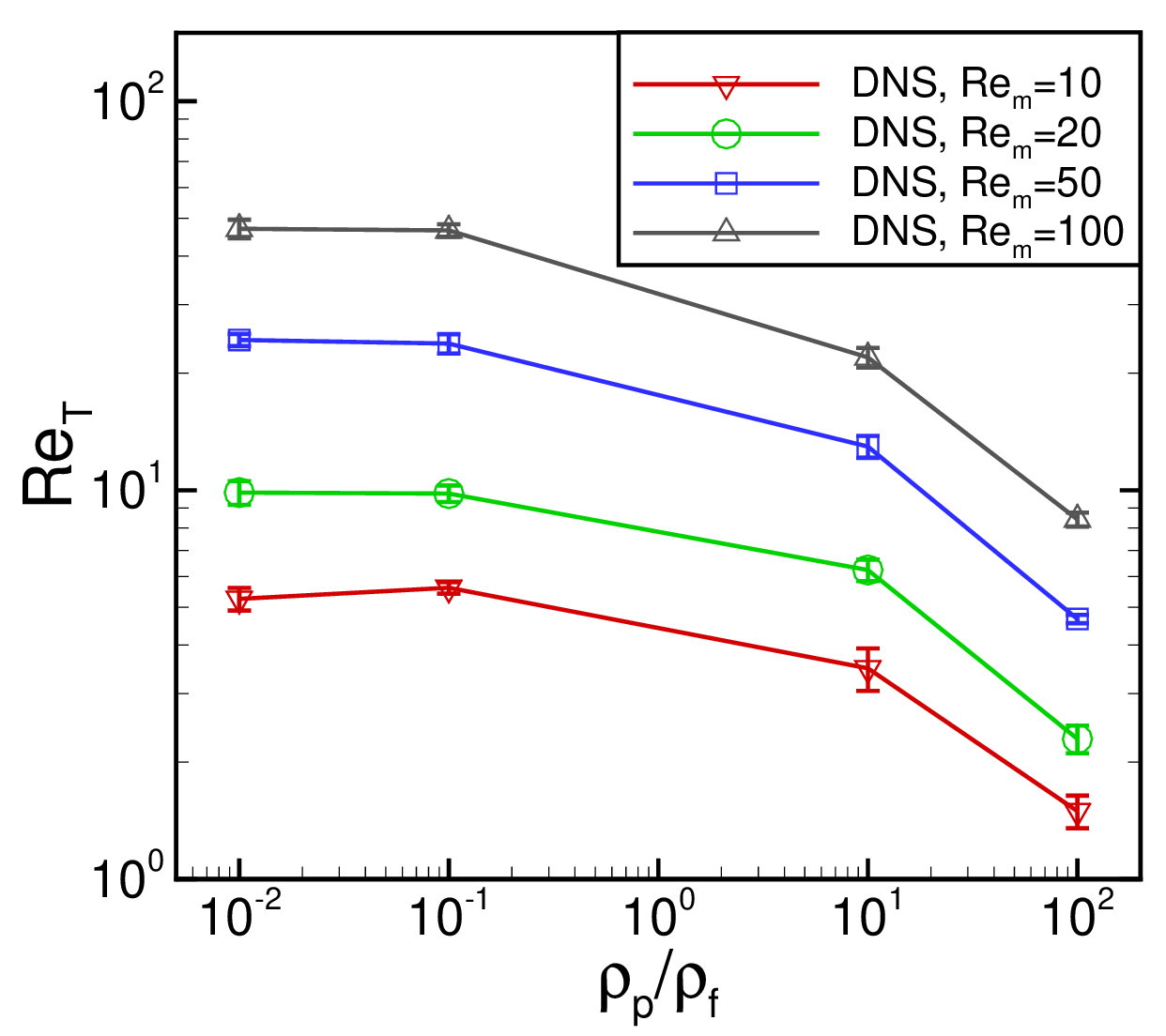}
\caption{Reynolds number based on the granular temperature $Re_T$, as a function of $\rho_p/\rho_f$ for different $Re_m$ and $\phi=0.4$.
The error bars represent $95 \%$ confidence intervals obtained from five independent realizations for each case.}
\label{fig:ReT}
\end{centering}
\end{figure}

Figure \ref{fig:ReT} shows the Reynolds number based on the granular temperature at the statistically stationary state for different $Re_m$ and $\rho_p/\rho_f$.
Previous studies have shown that the granular temperature increases continuously with decreasing density for heavy particles $\fn{\rho_p/\rho_f \geq 100}$ \citep{tenneti_jfm_2016,tang_free_2016,zaidi_2018}, and reaches an asymptotic value for buoyant particles \citep{tavana_acta_2019} which is the same trend seen in Fig. \ref{fig:ReT}.

\begin{figure} [H]
\begin{centering}
\subfigure[]{ \includegraphics[clip, width=60mm]{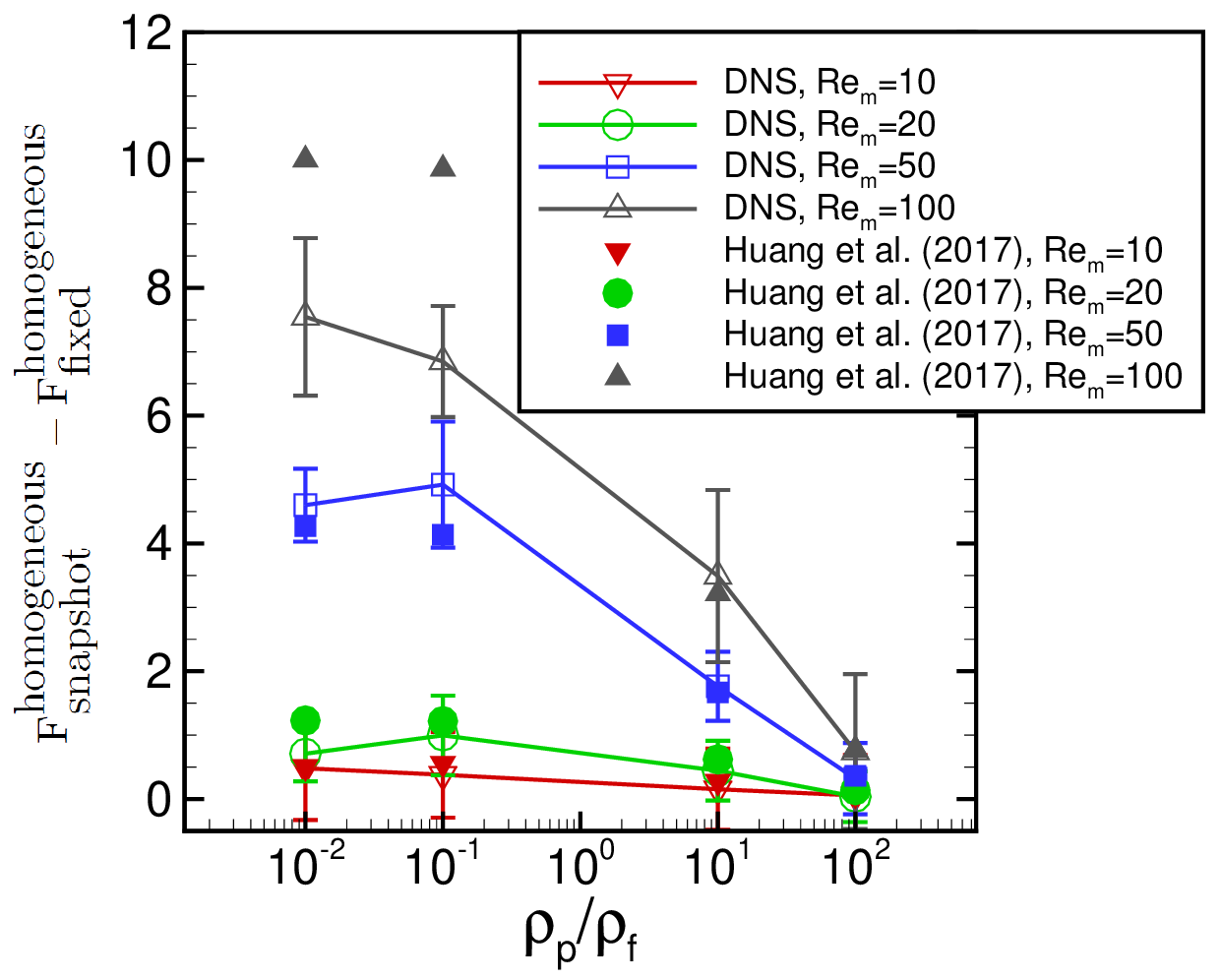} \label{fig:fluc_homo}}
\subfigure[]{ \includegraphics[clip, width=55mm]{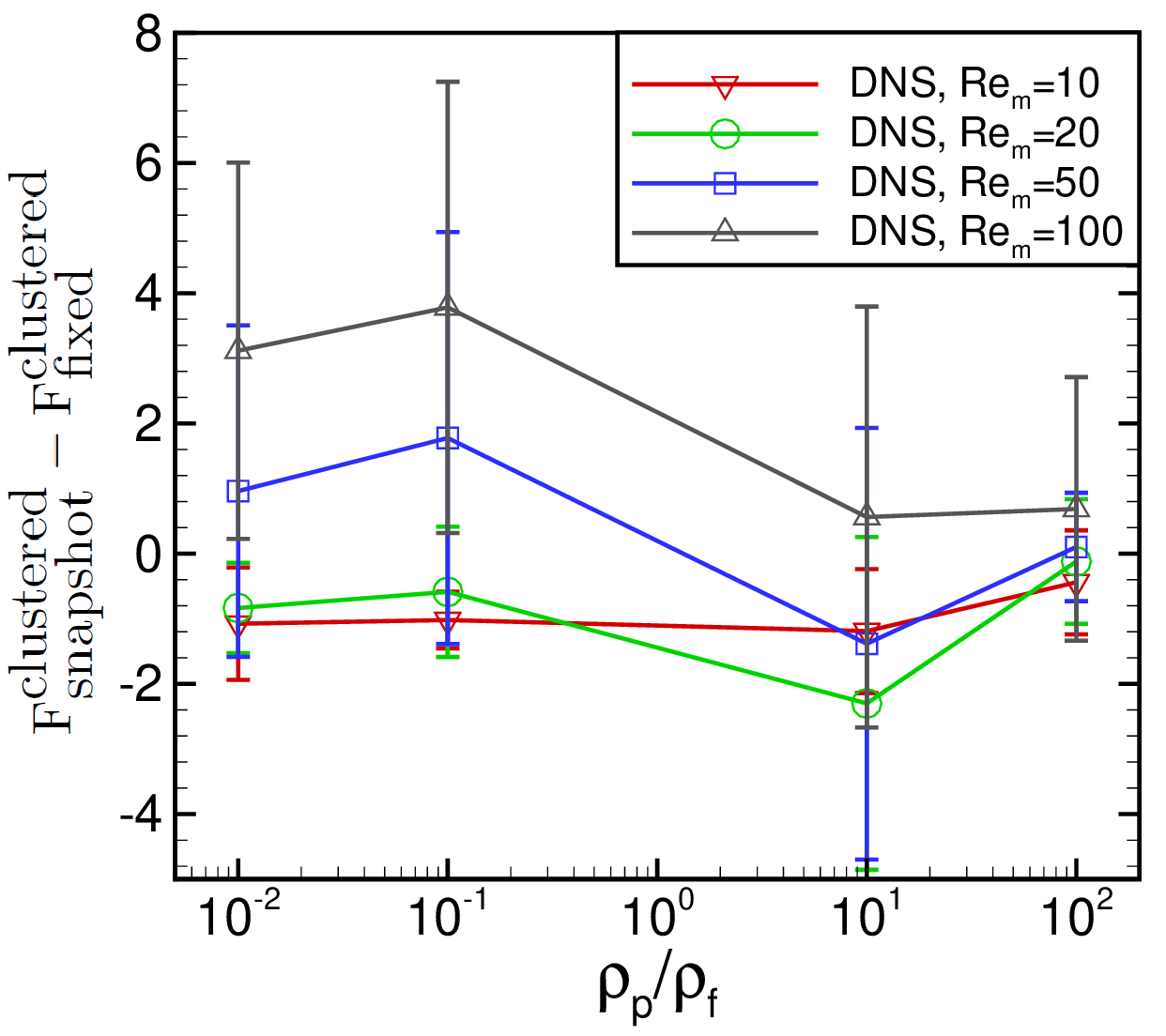} \label{fig:fluc_cluster}}
\caption{Drag difference scaled with Stokes drag between snapshot and fixed setups, as a function of $\rho_p/\rho_f$ for different $Re_m$ and $\phi=0.4$.
The error bars represent $95 \%$ confidence intervals obtained from five independent realizations for each case.
The corresponding $Re_T$ of each case in snapshot simulations is shown in Fig. \ref{fig:ReT}.
\subref{fig:rdf} Homogeneous (Cases 1 and 3).
The results from the correlation of \protect\citet{huang_2017} are also shown for the corresponding $Re_m$, $Re_T$, and $\phi$ for comparison.
\subref{fig:adf} Clustered (Cases 2 and 4).}
\label{fig:fluctuation}
\end{centering}
\end{figure}

To study the effect of particle velocity fluctuations on the drag experienced by a homogeneous configuration of particles, we compare the drag of the homogeneous snapshot setup (Case 3) with its homogeneous fixed counterpart (Case 1).
Figure \ref{fig:fluc_homo} shows an increase in drag for snapshot simulations when compared to homogeneous fixed beds.
The change in drag increases with decreasing density ratio and increasing Reynolds number.
This trend corresponds to an increase in particle velocity fluctuations or $Re_T$, according to Fig. \ref{fig:ReT}.
Figure \ref{fig:fluc_homo} also shows the increase of drag due to particle velocity fluctuations from the correlation of \citet{huang_2017} for the corresponding $Re_m$, $Re_T$, and $\phi$.
\citet{huang_2017} used a similar approach for developing their correlations.
However, they assigned a random fluctuating velocity to each particle in the fixed bed according to a Gaussian distribution corresponding to a specified value of the
particle granular temperature, while in our simulations, we get the velocity of each particle from the simulation of freely evolving suspensions at the statistically stationary state.
Huang's correlation is developed based on a dataset with $Re_T < 34.6$, and our results and their correlation match well in this range.
For higher values, their correlation overpredicts the increase in drag.
Note that we have calculated the value of Huang's correlation using the equivalent value of $Re_T$ from Fig. \ref{fig:ReT} for different density ratios.

Similar to the comparison of snapshot and fixed setup of homogeneous beds, we can compare the snapshot and fixed setup for clustered assemblies.
Figure \ref{fig:fluc_cluster} shows that unlike the homogeneous beds, the change of the drag for clustered snapshot simulations (Case 4) in comparison with clustered fixed beds (Case 2) does not have any specific trend with $\rho_p/\rho_f$ or $Re_m$.
This means that \textit{particle clustering affects the role that particle velocity fluctuations play in the drag force}.
In other words, particle clustering affects the drag force both directly (see Section \ref{sec:cluster}), and indirectly through particle velocity fluctuations.

\subsection{Effect of particle mobility}
\label{sec:mobility}
\begin{figure} [H]
\begin{centering}
\includegraphics[scale=0.4]{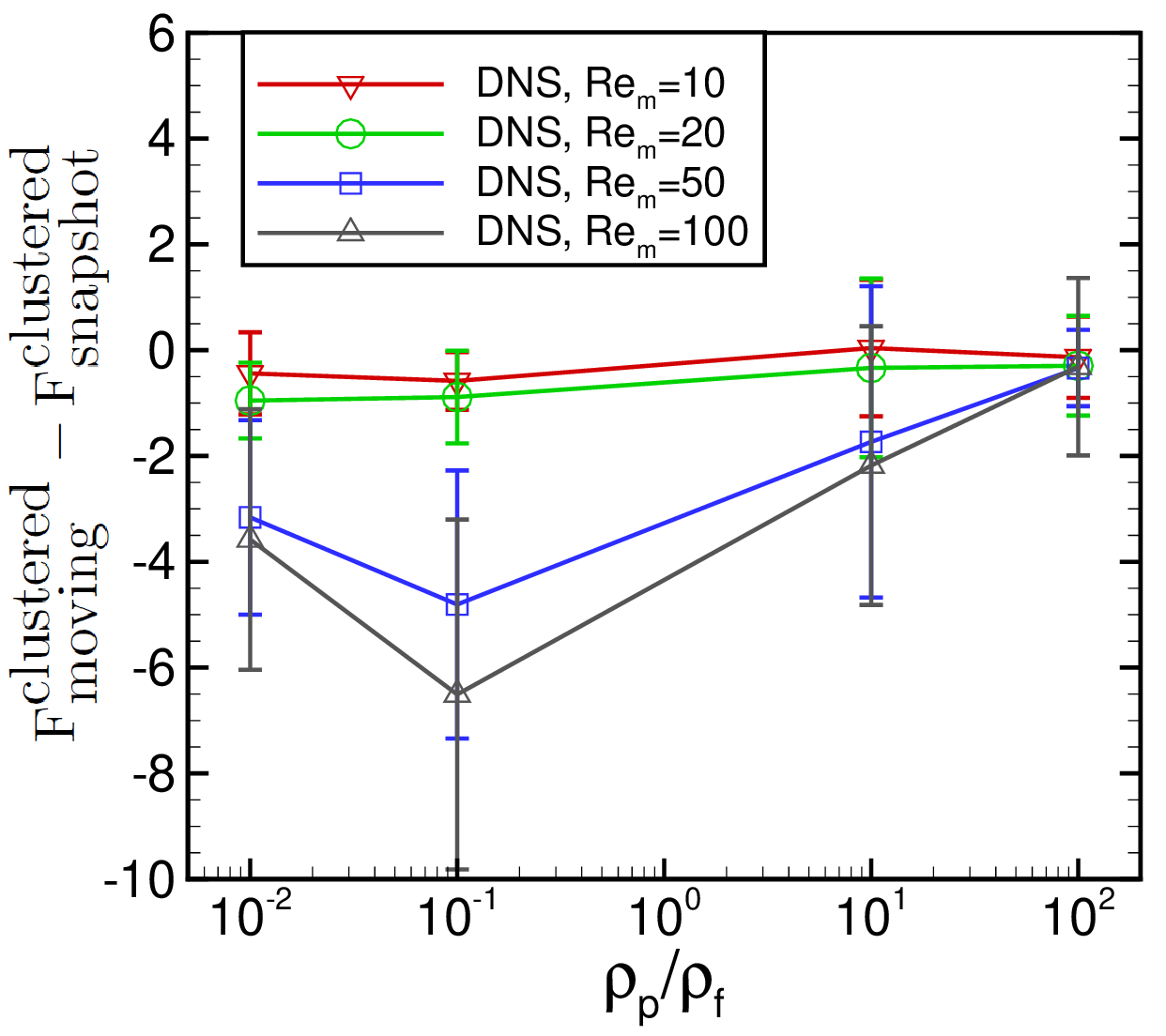}
\caption{Drag difference scaled with Stokes drag between freely evolving suspension and its snapshot counterpart (Cases 4 and 5), as a function of ${\rho_p/\rho_f}$ for different $Re_m$ and $\phi=0.4$.
The error bars represent $95 \%$ confidence intervals obtained from five independent realizations for each case.}
\label{fig:mobility}
\end{centering}
\end{figure}

To study the effect of particle mobility on drag, we have compared the drag of snapshot clustered simulations (Case 4) with freely evolving particles (Case 5).
Figure \ref{fig:mobility} shows a decrease in drag for moving particles when compared to snapshot clustered simulations.
The decrease is more significant for buoyant particles, since buoyant particles are more mobile and align their motion immediately with the streamlines of fluid flow.

One way to characterize the mobility of particles is by using the particle Stokes number, which is the ratio of the particle momentum response time to the characteristic time of the flow.
This number for a single solid particle in the Stokes flow regime is defined as:
\begin{equation}
\label{eq:St}
St =\frac{\rho_p \slip d_p}{18 \mu_f}. \nonumber
\end{equation}
For small Stokes number, the particles follow the streamlines of fluid, and for higher values, they continue moving on their initial trajectory.
To account for the effects of volume fraction, finite Reynolds number, and added mass, the Stokes number is modified as \citep{balachandar_review_2009}:
\begin{equation}
\label{eq:St_modified}
St = \frac{\fn{ {{\rho_p/\rho_f}}+C_{am}}}{18}\frac{Re_m}{\epf}\frac{1}{F_d\fn{Re_m,\phi}}.
\end{equation}
According to this definition, the Stokes number decreases with decreasing density ratio and reaches a constant value due to the added mass coefficient.

\subsection{Discussion on the range of density ratio considered in this study}
Before developing a drag law in the next section, it is useful to have a discussion on the physical relevance of the density ratio range $\fn{0.01 \le \rho_p/\rho_f \le 100}$ studied in this work.
This range covers some physical problems, for instance, microplastics, such as polyethylene and polypropylene, in the oceans $\fn{\rho_p / \rho_f \sim 0.9}$ \citep{driedger_2015}, different solid particles such as glass or sand in water $\fn{\rho_p / \rho_f \sim \mathcal{O} (1-10)} $, and cork in the air $\fn{\rho_p / \rho_f \sim 100} $ \citep{tenneti_jfm_2016}.
Obviously, there are some other applications of particle suspensions that are not covered here.
Although it might be useful to study a broader range of density ratios, it is computationally expensive.
In fact, the most significant variation in the behavior of the system due to the change of density ratio occurs in the range studied here.
Therefore the behavior of systems with a density ratio outside the range studied here can be predicted by the data presented in this work.

In fact, \citet{tavana_acta_2019} simulated particle suspensions with a wider range of density ratio including $\rho_p/\rho_f = 0.001$ and $1000$, but only for $Re_m=20$. 
Similar to the results presented here, they showed that the results are almost independent of the density ratio for buoyant particles.
This means that our conclusions about buoyant particles can be extended to lower density ratios, for example, air bubbles in contaminated water where $\rho_p / \rho_f \sim 0.001$.
There are also several industrial application of gas--solid flows with $\rho_p /\rho_f \geq \mathcal{O}(10^3)$.
\citet{tavana_acta_2019} also showed that the mean drag for $\rho_p / \rho_f = 1000$ is in good agreement with the mean drag in fixed beds.
Even our results for $\rho_p / \rho_f = 100$ are close to fixed beds.
In addition, it is well accepted that granular temperature for heavy particles is proportional to $\fn{\rho_p / \rho_f}^{-1}$ \citep{tenneti_jfm_2016,tang_free_2016}.
Overall, we expect that our conclusions for $\rho_p / \rho_f =100$ to be applicable to higher density ratios as well.

\section{A new drag law for freely evolving suspensions}
\label{sec:drag_law}
As mentioned in the Introduction, an accurate drag correlation is essential to perform predictive CFD simulations.
In this section, we propose a new drag law for interphase momentum transfer in particle suspensions based on the complete set of freely evolving simulations performed in this work.
To develop the correlation, we can follow two different strategies, which are discussed in this section.

\subsection{Drag law using symbolic regression}
In the first approach, we use symbolic regression to derive a correlation for mean drag.
To do this, we have used the HeuristicLab software package \citep{wagner_2014}.
The output of this software is a mathematical expression that fits the input data.
Alongside this expression, the model is also presented as a tree where each sub-tree is a term in the expression.
For each sub-tree, a \textit{node impact} is defined, which shows the importance of that sub-tree and has a value between zero and one (zero: not important, one: very important).
Therefore, the tree (and the corresponding mathematical expression) can be simplified by removing sub-trees that do not significantly affect the accuracy of the model.
Usually, the original expression provided by the software is long and complicated; therefore, it is necessary to simplify the model to come up with a simpler expression.
In developing our correlation, we have eliminated sub-trees which have a \textit{node impact} smaller than $0.001$.
In this software, we also have the option to choose the functional forms which can be used in developing the model such as polynominal power functions, exponential functions, and logarithmic functions.

To develop the correlation, we have to specify which variables the drag correlation should depend on.
In previous works, three different sets of variables have been used.
\citet{tang_free_2016} proposed a correlation in the form of 
${F}\fn{Re_m,\phi,Re_T}$.
\citet{rubinstein_2016} used ${F}\fn{\phi,St}$ format for low Reynolds flows, so their correlation does not depend on $Re_m$.
The correlation by \citet{zaidi_2018} is in the form of ${F}\fn{Re_m,\phi,\rho_p/\rho_f}$.
Note that the variables in these correlations are not independent: $Re_m \fn{\phi}$, $Re_T \fn{Re_m,\phi,\rho_p/\rho_f}$, and $St \fn{Re_m,\phi,\rho_p/\rho_f}$.

At the outset it is not clear which is the best choice for the set of independent variables. In general, assuming $\bld{x},\;\bld{y}$, and $\bld{z}$ represent the different independent variable spaces, we can formulate correlations as $f_1 \fn{x_1,x_2,x_3}$, $f_2 \fn{y_1,y_2,y_3}$, and $f_3 \fn{z_1,z_2,z_3}$, while recognizing that $\bld{y} = u \fn{\bld{x}}$ and $\bld{z} = w \fn{\bld{x}}$. 
We now examine which choice is the best and what are the performance metrics for comparing three different approaches.
To answer these questions, we have developed correlations in three spaces $\fn{Re_m,\phi,\rho_p/\rho_f}$, $\fn{Re_m,\phi,Re_T}$, and $\fn{Re_m,\phi,St}$ which already have been used in the literature.
It is also possible to define new variable spaces using data-driven dimensional analysis by measuring the relative importance of variables \citep{jofre_2020}, which is the subject of further studies.

The performance metrics we consider in this work to compare these three variable spaces are complexity, accuracy, and predictability.
We divide the input data randomly into $80 \%$ training and $20 \%$ test datasets.
The training dataset is used to measure the accuracy of the correlation by considering the average and maximum relative error between the correlations and data points.
Using the same criterion, the test dataset is used to measure the predictability of the correlation.
The complexity of each correlation is measured by considering the number of model constants in the correlation.

Table \ref{tab:compare_correlation} summarizes the features/performance of different correlations.
Although there is not a meaningful difference between the performance of the three correlations, the one which is a function of density ratio has a smaller error for both training and test datasets with fewer constants in the correlation.
This variable space also has the benefit that $\rho_p/\rho_f$ is an input parameter while $Re_T$ and $St$ are derived quantities.

\begin{table} [H]
\centering
\addtolength{\leftskip} {-3cm}
\caption{Comparison of correlations developed for drag law using different variable spaces.}
\label{tab:compare_correlation} 
\begin{tabular}{c c c c} 
\hline\noalign{\smallskip}
Parameter  & Complexity & Accuracy & Predictability \\
\noalign{\smallskip}
\hline\noalign{\smallskip}
Criterion  & \begin{tabular}{c} Number of constants \end{tabular} & \begin{tabular}{c} relative error of training dataset \\ Avg $\fn{\%}$ \;\;\;\;\;\;\; Max $\fn{\%}$\end{tabular} & \begin{tabular}{c} relative error of test dataset \\ Avg $\fn{\%}$ \;\;\;\;\;\;\; Max $\fn{\%}$\end{tabular} \\
\noalign{\smallskip}
\hline\noalign{\smallskip}
${F}\fn{Re_m,\phi,\rho_p/\rho_f}$ & $12$ & $4.41$ \;\;\;\;\;\;\; $18.26$ & $4.70$ \;\;\;\;\;\;\; $13.18$ \\ 
${F}\fn{Re_m,\phi,Re_T}$ & $12$ & $4.99$ \;\;\;\;\;\;\; $13.14$ & $6.74$ \;\;\;\;\;\;\; $17.05$ \\ 
${F}\fn{Re_m,\phi,St}$ & $19$ & $5.43$ \;\;\;\;\;\;\; $20.12$ & $5.50$\;\;\;\;\;\;\; $10.99$ \\ 
\hline\noalign{\smallskip}
\end{tabular}
\end{table}

Although using symbolic regression is straightforward, one may encounter some problem using this approach since we do not have so much control over the final functional form of the correlation.
For instance, the model developed in this approach is only valid in the range of input data used.
This is not a problem by itself, and of course, is to be expected.
However, the final goal of developing such correlations is to use them in large scale simulations such as EE or EL, which are used for systems with a very wide range of parameters.
So it is important that the proposed correlations at least show a reasonable trend when they are used out of their proposed range.
For the drag law of freely evolving suspensions, it is important that the proposed correlation converges to the drag of fixed beds for large values of density ratios, and goes to zero with decreasing volume fraction.
To overcome these problems, we included the data for fixed beds in our input dataset.
Besides, we added a few additional data points based on the physics to better satisfy the limiting cases such as zero drag at $\phi=0$.
By considering these modifications, our final correlation is (See \ref{sec:appB} for a corrigendum)
%
\begin{align}
\label{eq:Fd_rho}
{F}\fn{Re_m,\phi,\rho_p/\rho_f} = F_d / F_{st} = & \; c_1 + c_2 \; \phi + c_3 \; \phi  Re_m +  c_4 \; \phi^5 Re_m^{1/3} \nonumber \\
& +  \cfrac{ c_5 + Re_m + c_6 \; \phi Re_m^2 }{ c_7 + c_8 \; \rho_p / \rho_f }
\end{align}
with the following constants,
\begin{align}
\label{eq:constants}
c_1 & = 0.245, \;\; c_2 = 22.8, \;\;\;\; c_3 = 0.242, \;\;\;\;\; c_4 = 130.371, \nonumber \\
c_5 & = 6.708, \;\;c_6 = 0.233, \;\; c_7 = 140.272,\;\; c_8 = 2.299, \nonumber
\end{align}
which fits the data with a relative error less than $24 \%$.

\begin{figure} [H]
\begin{centering}
\subfigure[]{ \includegraphics[clip, width=57mm]{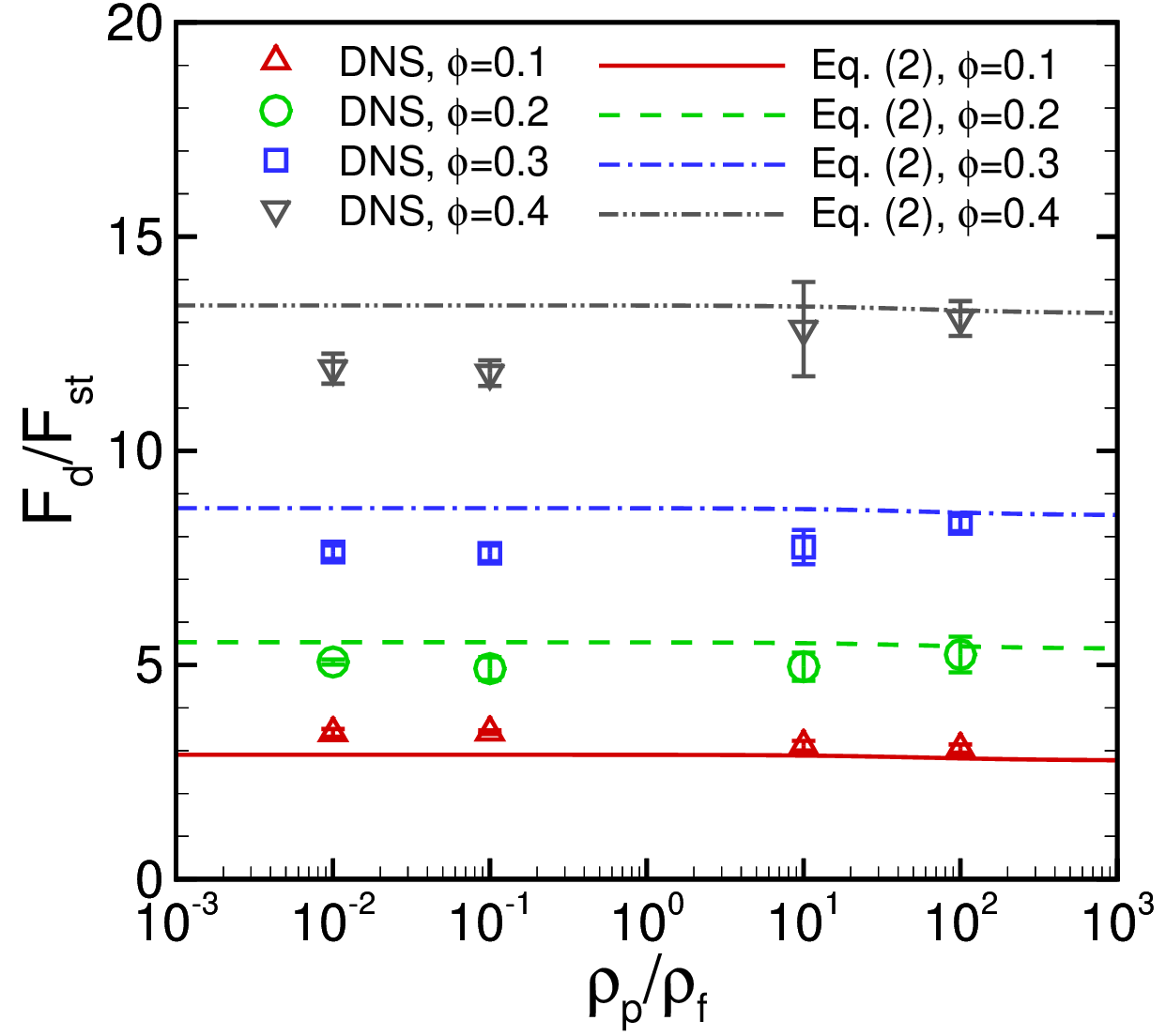} \label{fig:drag1_TPS}}
\subfigure[]{ \includegraphics[clip, width=57mm]{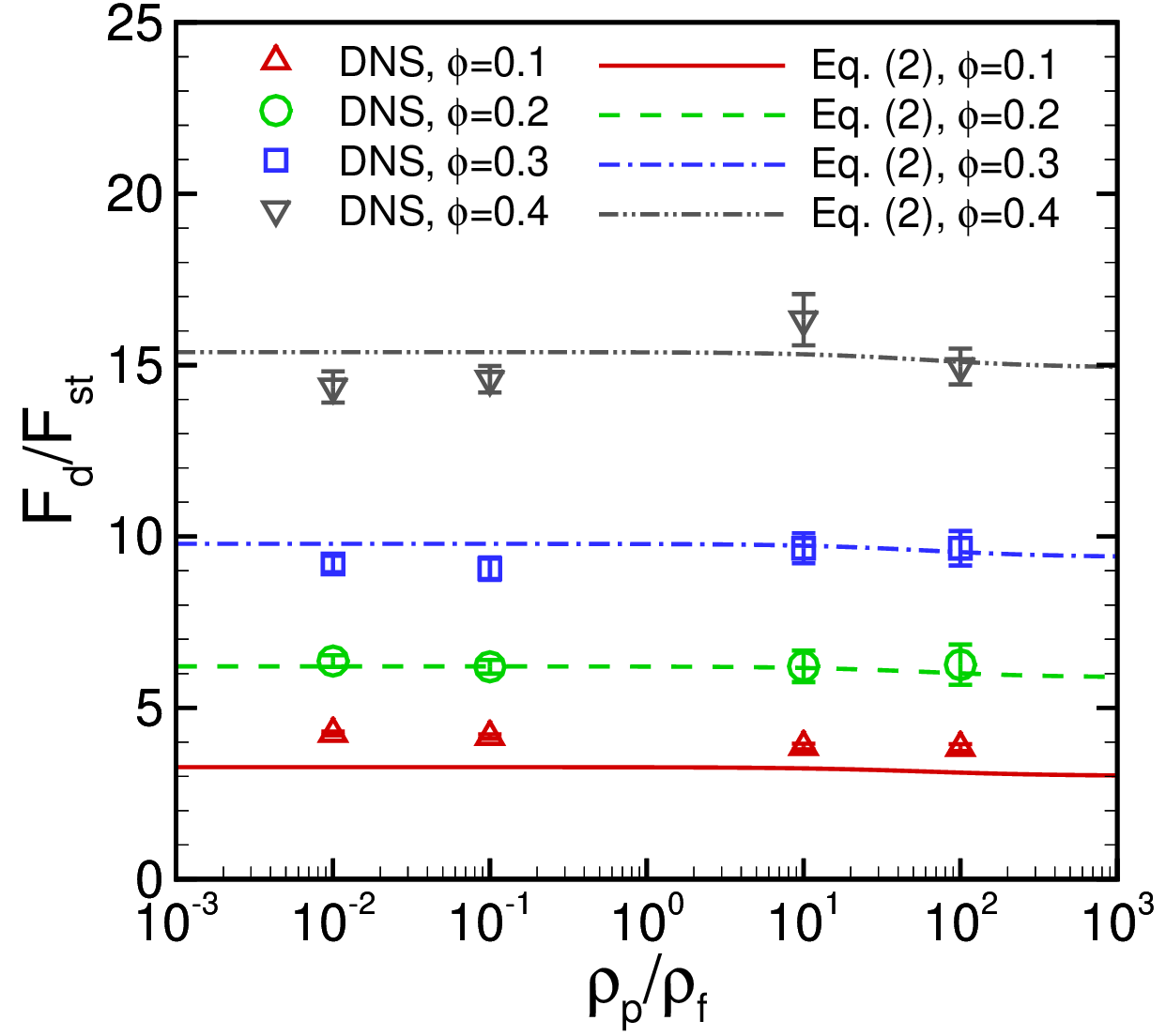} \label{fig:drag2_TPS}}
\subfigure[]{ \includegraphics[clip, width=57mm]{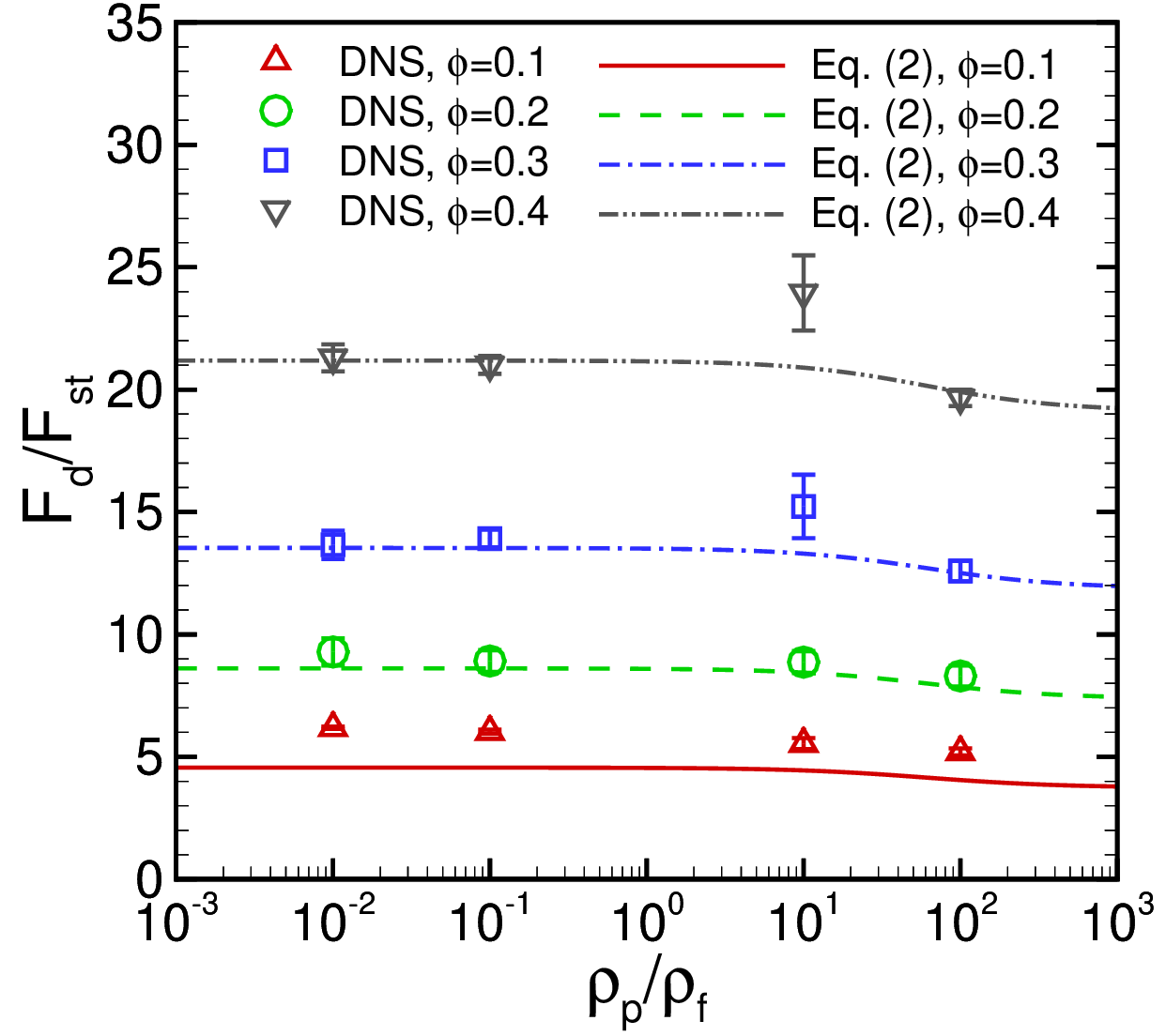} \label{fig:drag3_TPS}}
\subfigure[]{ \includegraphics[clip, width=57mm]{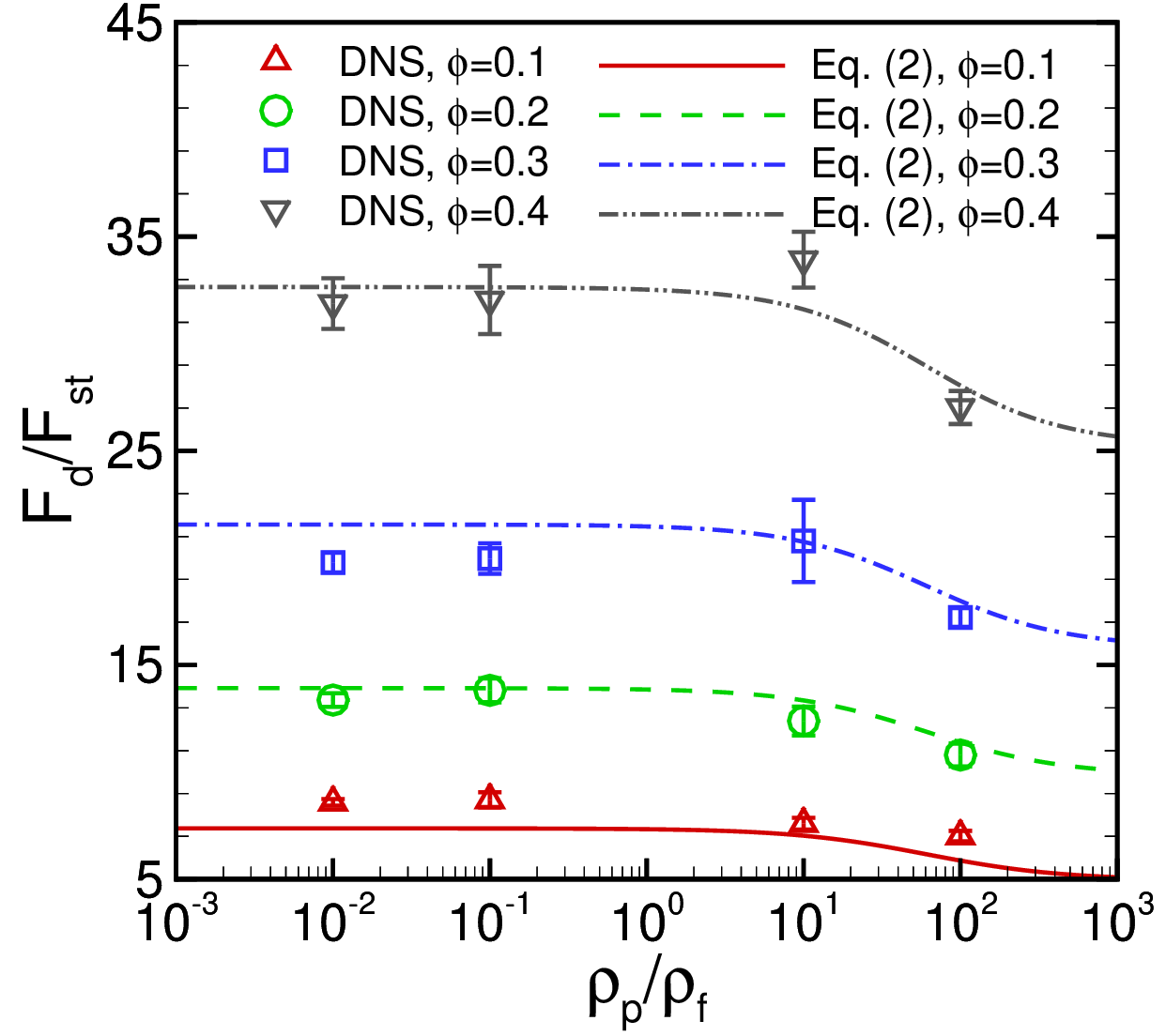} \label{fig:drag4_TPS}}
\caption{Non-dimensional drag force $F= F_d/F_{st}$, as a function of ${\rho_p/\rho_f}$ for different $Re_m$ and $\phi$.
The drag correlation by Eq. (\ref{eq:Fd_rho}) is also shown for comparison.
Symbols show PR-DNS data, and lines present the correlation.
The error bars represent $95 \%$ confidence intervals obtained from five independent realizations for each case.
\subref{fig:drag1_TPS} $Re_m=10$.
\subref{fig:drag2_TPS} $Re_m=20$.
\subref{fig:drag3_TPS} $Re_m=50$.
\subref{fig:drag4_TPS} $Re_m=100$.}
\label{fig:drag_TPS}
\end{centering}
\end{figure}

Figure \ref{fig:drag_TPS} shows the non-dimensional drag force $F = F_d/F_{st}$, as a function of density ratio for different Reynolds numbers and volume fractions from PR-DNS and the new drag correlation, i.e. Eq. (\ref{eq:Fd_rho}), for comparison.

Before finishing this subsection, we want to mention other problems with symbolic regression, which one may encounter.
It is possible that the developed correlation is singular for specific combinations of $Re_m$, $\phi$, and $\rho_p / \rho_f$ if this results in a zero value in the denominator.
Our correlation becomes singular for a negative value of density ratio, which is not a problem.
Another problem with symbolic regression is that the results are not reproducible, which means that every time we run the software, we may come up with a different model.
This is because of the use of genetic algorithm, which is an stochastic search algorithm, in symbolic regression.
Considering all these possible difficulties/problems with symbolic regressions, we discuss proposing correlations using a predefined functional form in the next subsection.

\subsection{Drag law using predefined functional forms}
The second approach for developing drag correlations is using a predefined functional form for the correlation, which is informed by the physics of the problem.
Generally, the correlations used to describe the drag force in general flow conditions are classified into two forms.
In the first approach, the correlation is based on the drag force in the limit of Stokes flow regime to which a term linear in $Re_m$ is added accounting for the inertial effects,
\begin{equation}
\label{eq:drag_1st}
F_d\fn{Re_m,\phi} = F_d \fn{0,\phi} + \alpha Re_m.
\end{equation}
Originally, $\alpha$ was only a function of volume fraction \citep{ergun_1952}.
However, it was later shown that $\alpha$ also depends on $Re_m$ \citep{hill_koch_2001b,beetstra_2007,tenneti_ijmf_2011}.
Recently, \citet{tang_free_2016} proposed a drag law for moving particles by adding a term to their fixed bed correlation \citep{tang_2015}.
In other words, they proposed a correlation for the change of drag in moving particles in comparison to its fixed bed counterpart.

Although it seems more convenient to use Tang's form of the correlation for incorporating the particle motion in fixed bed drag laws, we could not find a simple and accurate model for $\Delta F$ for our dataset using symbolic regression.

In the second form of drag laws, the relation is based on the drag force on a single particle, where the influence of the other particles is accounted for by multiplying with a power of the voidage,
\begin{equation}
\label{eq:drag_2nd}
F_d\fn{Re_m,\phi} = F_d \fn{Re_m,0} \epf^{-n}.
\end{equation}
The value of $n$ was originally constant \citep{wen_yu_1966} but it was later shown that it is also a function of Reynolds number \citep{difelice_1994,rong_2013}.
Most recently, \citet{zaidi_2018} showed that the dependence of $n$ on density ratio should also be considered for heavy moving particles.

To utilize this form of the drag law, we use the Schiller--Naumann drag law for $F_d \fn{Re_m,0}$, i.e.,
\begin{equation}
\label{eq:scaled_drag_Fsingle}
F_d\fn{Re_m,0} = F_{single} = F_{st} \fn{ 1+0.15 Re_m^{0.687} }.
\end{equation}

\begin{figure} [H]
\begin{centering}
\subfigure[]{ \includegraphics[clip, width=57mm]{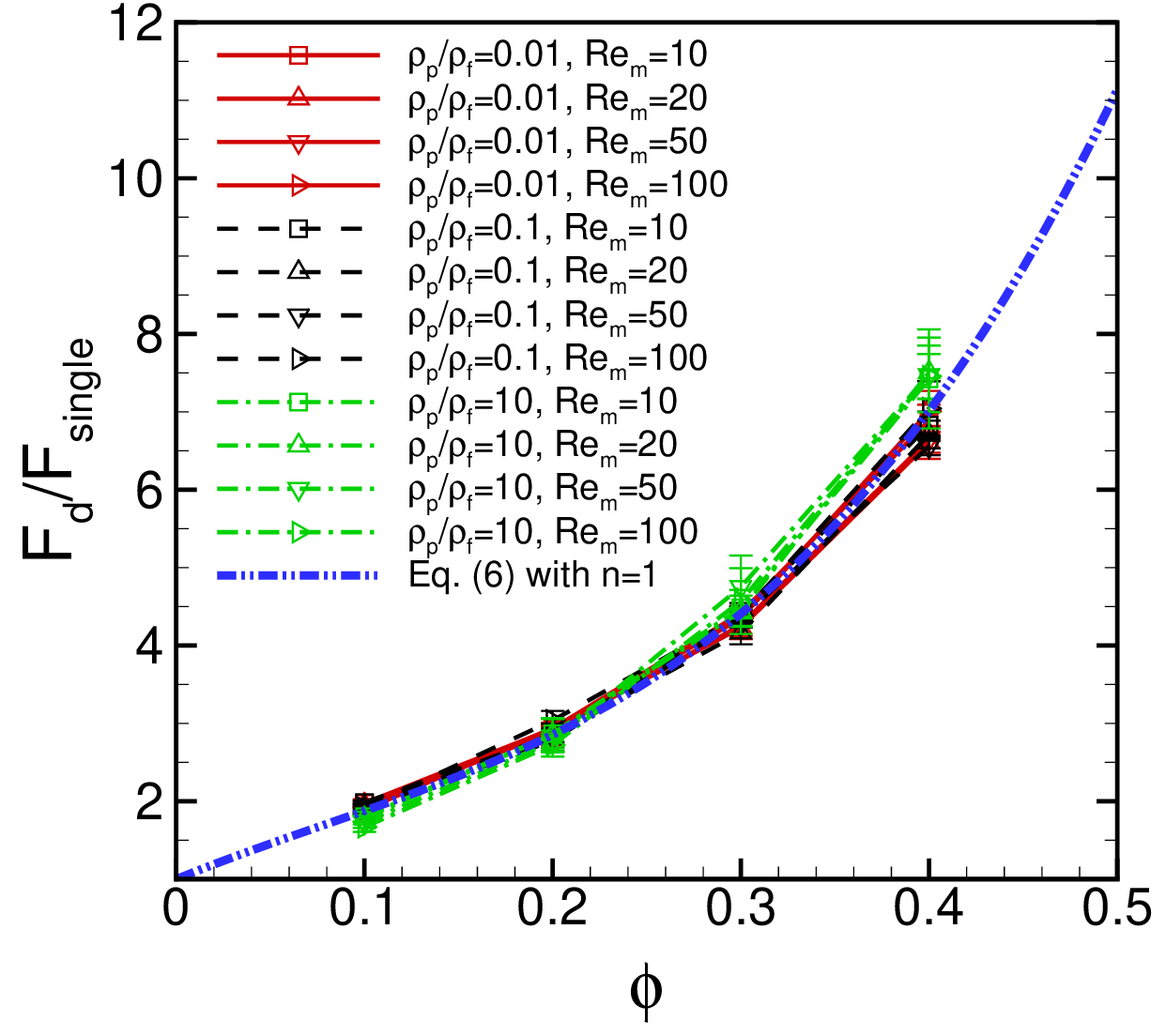} \label{fig:drag_Fsingle1}}
\subfigure[]{ \includegraphics[clip, width=57mm]{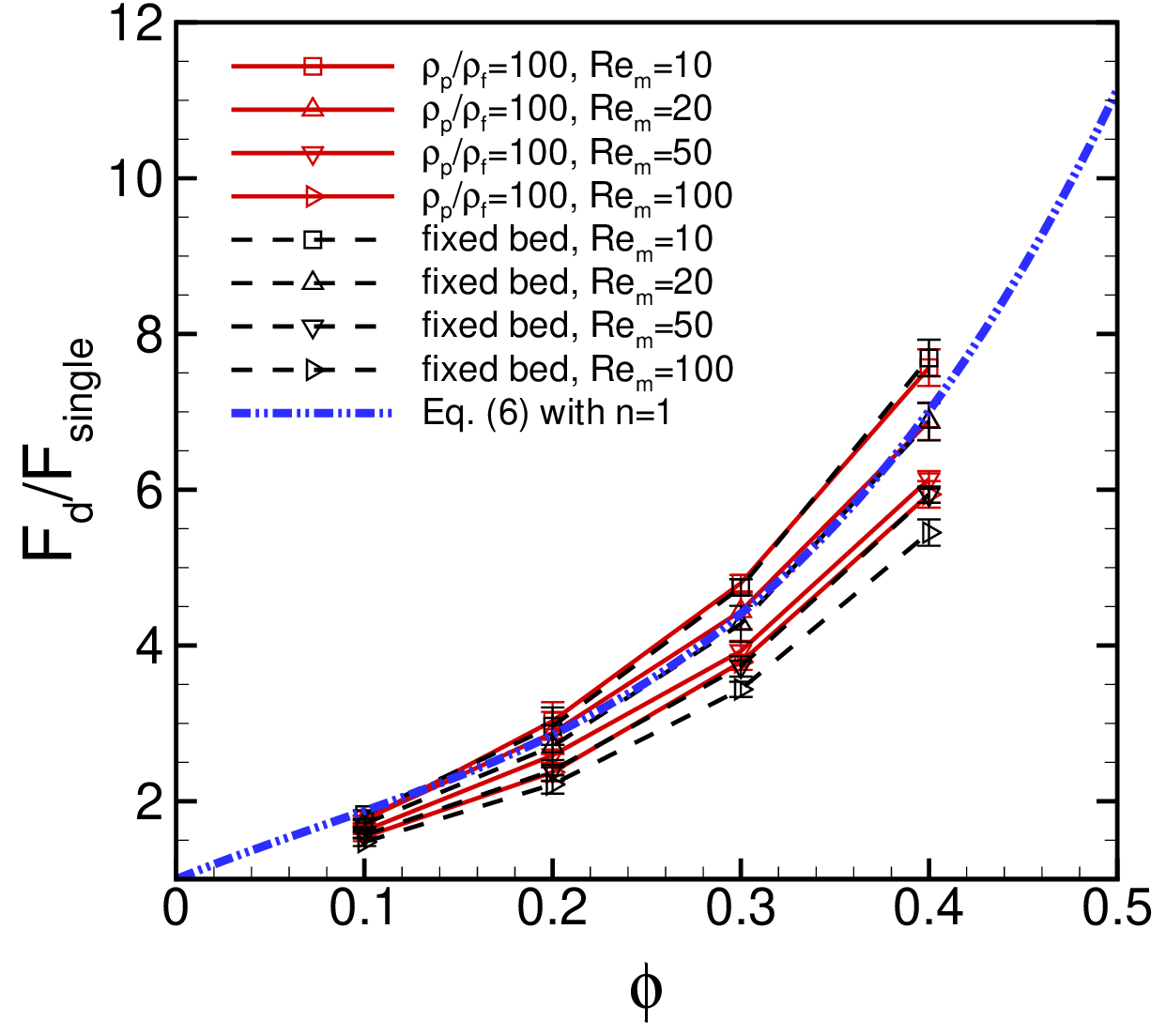} \label{fig:drag_Fsingle2}}
\caption{Non-dimensional drag force $F_d/F_{single}$, as a function of $\phi$ for different $Re_m$ and $\rho_p / \rho_f$.
The drag correlation by Eq. (\ref{eq:drag_Fsingle_correlation}) is also shown for comparison.
Symbols show PR-DNS data, and lines present the correlation.
The error bars represent $95 \%$ confidence intervals obtained from five independent realizations for each case.
\subref{fig:drag_Fsingle1} $\rho_p / \rho_f = 0.01, 0.1, 10$.
\subref{fig:drag_Fsingle2} $\rho_p / \rho_f = 100$ and fixed beds.}
\label{fig:drag_Fsingle}
\end{centering}
\end{figure}

The results with this new scaling are presented in Fig. \ref{fig:drag_Fsingle}.
Interestingly, the results for all Reynolds number and $\rho_p / \rho_f \le 10$ approximately collapse to a single line. 
This means for this range of parameters; the new scaled drag is only a function of volume fraction.
However, the new scaled drag for $\rho_p / \rho_f = 100$ shows a dependence on Reynolds number in addition to $\phi$ (see Fig. \ref{fig:drag_Fsingle2}).
In Fig. \ref{fig:drag_Fsingle2}, the drag for fixed beds with this new scaling is also shown.
It is clear that the drag of fixed beds is close to the drag of moving particles for $\rho_p / \rho_f = 100$.
By curve fitting using MATLAB, and using the fact that $F_d/F_{single}$ should be unity at $\phi=0$, we will have $\left( R^2 = 0.9852 \right)$:
\begin{equation}
\label{eq:drag_Fsingle_correlation}
\frac{F_d}{F_{single}} = \fn{78.96 \phi^3 - 18.63 \phi^2 + 9.845 \phi + 1}^n.
\end{equation}
where $n=1$ for light particles with $\rho_p / \rho_f \le 10$ (liquid--solid and bubbly flows).
To improve this correlation for gas--solid flows $\rho_p / \rho_f = 100$, we propose the following expression for $n$ which is inspired by \citet{garside_1977},
\begin{equation}
\label{eq:n_value}
\frac{1.05 - n}{n - 0.9} = 4.3 \times 10^{-4} Re_m^{2.361}.
\end{equation}
The correlation given by Eqs. (\ref{eq:drag_Fsingle_correlation}) and (\ref{eq:n_value}) fits the data with a relative error less than $14 \%$ which is an improvement over the correlation presented by Eq. (\ref{eq:Fd_rho}).
Also note that comparing Eqs. (\ref{eq:drag_2nd}) and (\ref{eq:drag_Fsingle_correlation}), we have used a third degree polynomial instead of $\epf^{-1}$ to get a better fit.

Generally, the drag law is used to model the unclosed average interphase momentum transfer term in the mean momentum conservation equation of the two-fluid theory, and determines the overall mean particle-laden flow structure. 
Since our drag law is inferred from freely evolving particle suspensions, the effect of the motion of the particles is captured in the drag correlation. 
This improved drag law can enhance the predictive capability of CFD simulations of particle-laden flows that are based on the two-fluid theory. 
The improved drag law can also be used to refine the stability limits for particle-laden suspensions since these limits are determined by the functional dependence of drag on volume fraction.

Before concluding our work, we want to emphasize two points.
First, our correlations apply to homogeneous flows.
Although we discussed the clustering of particles in Section \ref{sec:cluster}, this clustering, as we mentioned earlier, only happens at small scales (less than $1.5 d_p$) and we call it \textit{short-range clustering}, in contrast to, long-range clustering in gas--solid flows where the length scale of clusters is $O \fn{10 - 100} d_p$ \citep{capecelatro_2015}. 
This means that particles are dispersed approximately homogeneously in the domain at large scales, which is consistent with previous works on bubbly flows \citep{bunner_tryggvason_2002a,bunner_tryggvason_2003}.
Second, the focus of this work is on the effect of motion of particles on the drag compared to fixed beds.
However, another important topic in particle-laden flows is the distribution of force on different particles \citep{akiki_2016,huang_2017,esteghamatian_2018} which is an essential consideration in developing stochastic \citep{tenneti_jfm_2016, esteghamatian_2018,lattanzi_JFM_2020} or deterministic \citep{akiki_jackson_balachandar_2017,akiki_2017,seyedahmadi_wachs_2020} models for point-particle simulations.

\section{Conclusions}
\label{sec:conclusion}
PR-DNS are performed for a wide range of Reynolds number $\fn{10 \le Re_m \le 100}$, volume fraction $\fn{0.1 \le \phi \le 0.4}$, and density ratio $\fn{0.01 \le \rho_p/\rho_f \le 100}$.
The effects of particle clustering, particle velocity fluctuations, and mobility of particles on the interphase momentum transfer of dispersed multiphase flows are studied.
It is shown that clustering in the particle configuration increases the drag by generating a horizontal raft.
Although particle velocity fluctuations characterized by the Reynolds number based on granular temperature increase the drag for homogeneous configurations, it is shown that the effects of particle clustering and particle velocity fluctuations are not independent.
Overall, the combined effects of particle clustering and particle velocity fluctuations decrease or increase the drag depending on flow conditions.
It is also found that the mobility of particles decreases the drag.
Finally, it is shown that the competing effects of these factors could result in an increase, decrease, or no change of drag in freely evolving suspensions in comparison to fixed beds.

A new drag law for monodisperse suspensions is proposed using two different approaches: symbolic regression and predefined functional forms. 
In the symbolic approach, the correlation is proposed as a function of $F \fn{Re_m, \phi, \rho_p/\rho_f}$ to incure the least error for both training and test datasets with a fewer number of constants in the correlation in comparison with other variable spaces.
It is also shown that we can develop a correlation using predefined functional forms which only depend on volume fraction for $ \rho_p / \rho_f \le 10$.

\section*{Acknowledgment}
This material is based upon work supported by the National Science Foundation under Grant no. CBET 1438143.

\begin{appendix}
\section{Drag force and total fluid-particle drag}
\label{sec:appA}
At the statistically stationary state, each particle experiences two forces from the fluid, a body force $\bld{F}_{mpg}$ due to mean pressure gradient and a force $\bld{F}_d$ resulting from the fluctuating pressure field and the viscous contribution to the drag force. 
The sum of these two forces is the total fluid-particle force $\bld{F}_{g \rightarrow s}$ that the fluid exerts on a particle. 
The forces are related to the pressure drop over the system as follow:
\begin{equation}
\label{eq:balance_lab1}
	- \left\langle \boldsymbol{\nabla}P \right\rangle_v = \frac{N_p}{\volume_{sys}} \bld{F}_{g \rightarrow s} = \frac{N_p}{\volume_{sys}} \fn{ \bld{F}_d - V_p \left\langle \boldsymbol{\nabla}P \right\rangle_v},
\end{equation}
where $\volume_p$ is the volume of a single particle. 
Eq. \ref{eq:balance_lab1} can be written as:
\begin{equation}
\label{eq:balance_lab2}
	- \left\langle \boldsymbol{\nabla}P \right\rangle_v = \frac{\phi}{1 - \phi} \frac{\bld{F}_d}{\volume_{p}}.
\end{equation}
From Eqs. (\ref{eq:balance_lab1}) and (\ref{eq:balance_lab2}), it follows that,
\begin{align}
\label{eq:relation_lab}
	\bld{F}_{g \rightarrow s} & = \bld{F}_{d} + \bld{F}_{mpg} = \bld{F}_{d} - \left\langle \boldsymbol{\nabla}P \right\rangle_v \volume_p, \nonumber \\
	\bld{F}_{mpg} & = \frac{\phi}{1 - \phi} \bld{F}_d, \nonumber \\
	\bld{F}_{g \rightarrow s} & = \frac{\bld{F}_{d}}{1-\phi} = \frac{\bld{F}_{mpg}}{\phi}.
\end{align}

In PR-DNS, $\bld{F}_d$ and $\bld{F}_{g \rightarrow s}$ are calculated by averaging over all the particles.
In the literature, both $\bld{F}_d$ \citep{vanderHoef_2005,beetstra_2007,tavana_acta_2019} and $\bld{F}_{g \rightarrow s}$ \citep{hill_koch_2001a,hill_koch_2001b,tenneti_ijmf_2011,tang_2015} are reported for proposing drag laws, and their relation is $\fn{1-\phi} \bld{F}_{g \rightarrow s} = \bld{F}_d$ as shown in Eq. (\ref{eq:relation_lab}).
In this work, we have reported $\abs{\bld{F}_d}$.

\section{Corrigendum}
\label{sec:appB}
The authors would like to draw attention to the fact that the drag correlation presented in Section 5.1 by Eq. (2) converges to almost zero for the limit of zero volume fraction.
However, it should converge to the drag on a single particle in this limit.
To correct this trend, we define the $p-$norm of the drag on an isolated particle $f_{iso}$ and the proposed correlation in the article for finite volume fractions $f_{\phi}$ as the modified drag correlation,
\begin{align}
\label{eq:Fd_rho_correct}
F\fn{Re_m,\phi,\rho_p/\rho_f} = {F_d}/F_{st} = \fn{f_{iso}^p + f_{\phi}^p}^{\fn{1/p}}
\end{align}
where $ f_{iso} = \fn{1 + 0.15 Re_m^{0.687}} $ is calculated from the well-known Schiller--Naumann drag law and $f_{\phi}$ is the drag correlation given in the original paper for finite volume fractions, i.e., Eq. (2), which is repeated here
\begin{align}
f_{\phi} = & \; c_1 + c_2 \; \phi + c_3 \; \phi  Re_m +  c_4 \; \phi^5 Re_m^{1/3} \nonumber \\
& +  \cfrac{ c_5 + Re_m + c_6 \; \phi Re_m^2 }{ c_7 + c_8 \; \rho_p / \rho_f } \nonumber
\end{align}
with the following constants,
\begin{align}
c_1 & = 0.245, \;\; c_2 = 22.8, \;\;\;\; c_3 = 0.242, \;\;\;\;\; c_4 = 130.371, \nonumber \\
c_5 & = 6.708, \;\;c_6 = 0.233, \;\; c_7 = 140.272,\;\; c_8 = 2.299. \nonumber
\end{align}
With a proper choice of $p$ in Eq.~\ref{eq:Fd_rho_correct}, the $p-$norm will converges to $f_{iso}$ at low volume fraction since $f_{iso} > f_{\phi}$ for small volume fractions and it converges to $f_{\phi}$ for $\phi \ge 0.1$ since $f_{\phi} > f_{iso}$ in this range. 
Our numerical experiments show that the desired behavior is observed for $p=5$.
As indicated, this new modified correlation is technically the same as the original correlation for $\phi \ge 0.1$; therefore, the figures in the article do not change.

\begin{figure} [H]
\begin{centering}
\includegraphics[clip, width=80mm]{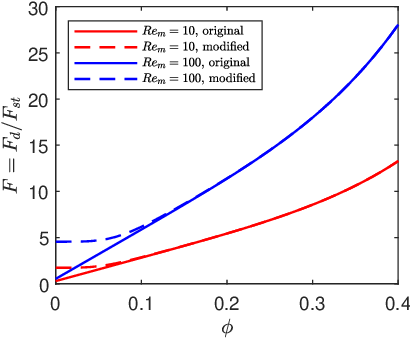}
\caption{The comparison between the modified drag correlation (Eq. (\ref{eq:Fd_rho_correct})) with $p=5$ and the original one (Eq. (\ref{eq:Fd_rho})) for ${\rho_p/\rho_f = 0.01}$ and $Re_m=10$ and $100$.}
\label{fig:drag_TPS_corrected}
\end{centering}
\end{figure}

For the sake of completeness, we present Fig.~\ref{fig:drag_TPS_corrected} here which compares the behavior of the original and modified correlations as a function of volume fraction for ${\rho_p/\rho_f = 0.01}$ and $Re_m=10$ and $100$.

\end{appendix}

\bibliography{mybibfile}

\end{document}